\pdfoutput=1
\documentclass[11pt,reqno]{amsart}
\usepackage{a4wide}
\usepackage{amssymb,amsmath,amsthm,enumerate,amsfonts}
\usepackage{newlfont}
\usepackage{dsfont}
\usepackage{amsbsy}
\usepackage[dvips]{graphicx}
\usepackage{verbatim} 
\usepackage{float}
\usepackage{cite}
\usepackage{color}

\usepackage{tikz}
\usepackage{booktabs}

\usepackage{url}
\usepackage{marginnote}

\usepackage{mathtools}
\usepackage{bbding, pifont}

\newcommand{\xmark}{\ding{55}}

\newcommand{\checkno}{\xmark}

\newcommand{\abs}[1]{\left|#1\right|}
\newcommand\norm[1]{\left\lVert#1\right\rVert}

\newcommand{\figdir}{.}

\def\hcboxcm#1#2{\hbox to #1{\hfill #2 \hfill}}

\def\null{\hbox{}}
\let\eps\varepsilon

\def\tn1{\widetilde n_1}
\def\tn2{\widetilde n_2}
\def\tn{\widetilde n }

\let\ds\displaystyle

\def\be{\begin{equation}}
\def\ee{\end{equation}}
\def\bea{\begin{eqnarray}}
\def\eea{\end{eqnarray}}

\def\bean{\begin{eqnarray*}}
\def\eean{\end{eqnarray*}}

\def\={\, = \, }

 \def\OO{\rm \hbox{O\kern-.34em\raise.47ex
         \hbox{$\scriptscriptstyle |$}\kern-.46em\raise.47ex
         \hbox{$\scriptscriptstyle |$}\kern+0.5 em }}
\def\RR{{\mathbb R} }

\def\OO{\mathbb{O}}

\def\RR{\mathbb{R}}

\def\Box{\leavevmode\vbox{\hrule
     \hbox{\vrule\kern4pt\vbox{\kern4pt}%
           \vrule}\hrule}}
\def\blackbox{\leavevmode\vrule height 5pt width 4pt depth 0pt\relax}
\catcode`@=11

\def\eqalign#1{\null\,\vcenter{\openup1\jot \m@th
   \ialign{\strut \hfil$\displaystyle{##}$ & $\displaystyle{{}##}$\hfil
      \crcr#1\crcr}}\,}
\def\eqalignrll#1{\null\,\vcenter{\openup1\jot \m@th
   \ialign{\strut \hfil$\displaystyle{##}$ & $\displaystyle{{}##}$\hfil
    & $\displaystyle{{}##}$\hfil
      \crcr#1\crcr}}\,}
\def\eqalignrcl#1{\null\,\vcenter{\openup1\jot \m@th
   \ialign{\strut \hfil$\displaystyle{##}$ &\hfil $\displaystyle{{}##}$\hfil
    & $\displaystyle{{}##}$\hfil
      \crcr#1\crcr}}\,}
\def\eqalignno#1{\displ@y \tabskip\@centering
  \halign to\displaywidth{\hfil$\@lign\displaystyle{##}$\tabskip\z@skip
    &$\@lign\displaystyle{{}##}$\hfil\tabskip\@centering
    &\llap{$\@lign##$}\tabskip\z@skip\crcr
    #1\crcr}}
\newcounter{appendix}
\newcounter{sectionz}
\def\appendix{\advance\c@appendix by 1
\def\thesectionz {\Alph{appendix}}
\def\thesection{\Alph{appendix}} 
   \ifnum\c@appendix=1 \setcounter{section}{-1} \fi
   \@startsection {section}{1}{\z@}{-3.5ex plus -1ex minus 
  -.2ex}{2.3ex plus .2ex}{\large\bf}}

\catcode`@=12
\newtheorem{lemme}{Lemma}[section]  

\newtheorem{theorem}[lemme]{Theorem}

\newtheorem{corollary}[lemme]{Corollary}

\newtheorem{definition}[lemme]{Definition}

\newtheorem{proposition}[lemme]{Proposition}

\newtheorem{remark}[lemme]{Remark} 

\def\deblem{\begin{lemme}\it }
\def\finlem{\end{lemme}}
\def\debthm{\begin{theorem}\it }
\def\finthm{\end{theorem}}
\def\debprop{\begin{proposition} \it}
\def\finprop{\end{proposition}}
\def\debcor{\begin{corollary}\it }
\def\fincor{\end{corollary}}
\def\debdef{\begin{definition}\it}
\def\findef{\end{definition}}
\def\debrem{\begin{remark}\em}
\def\finrem{\null\hfill\blackbox\end{remark}}
\newcommand{\vv}[1]{\textquotedblleft #1\textquotedblright }

\newcommand \SPeta {$\left(SP\right)_\eta$}
\newcommand \APeta {$\left(AP\right)_\eta$}

\title[AP-scheme for the vorticity equation]
{Asymptotic-Preserving scheme for a strongly anisotropic vorticity equation arising in fusion plasma modelling}

\author[A. Mentrelli, C. Negulescu]{Andrea Mentrelli$^\dagger$, Claudia Negulescu$^*$}
\address{$\dagger$ Department of Mathematics \& Alma Mater Research Center on Applied Mathematics (AM$^2$), University of Bologna, Italy;}
\address{$^*$ Universit\'e de Toulouse \& CNRS, UPS, Institut de Math\'ematiques de Toulouse UMR 5219, F-31062 Toulouse, France}

\email{andrea.mentrelli@unibo.it; claudia.negulescu@math.univ-toulouse.fr}

\date{\today}

\begin{document}

\maketitle

\begin{abstract}
The electric potential is an essential quantity for the confinement process of tokamak plasmas, with important impact on the performances of fusion reactors. Understanding its evolution in the peripheral region -- the part of the plasma interacting with the wall of the device -- is of crucial importance, since it governs the boundary conditions for the burning core plasma. The aim of the present paper is to study numerically the evolution of the electric potential in this peripheral plasma region. In particular, we are interested in introducing an efficient Asymptotic-Preserving numerical scheme capable to cope with the strong anisotropy of the problem as well as the non-linear boundary conditions, and this with no huge computational costs. This work constitutes the numerical follow-up of the more mathematical paper by C. Negulescu, A. Nouri, Ph. Ghendrih, Y. Sarazin, {\it Existence and uniqueness of the electric potential profile in the edge of tokamak plasmas when constrained by the plasma-wall boundary physics}. 
\end{abstract}

\bigskip


\maketitle

\section{Introduction} \label{SEC1}

The subject matter of the present paper is related to the magnetically confined fusion plasmas with the objective to contribute by some means to the improvement of the numerical schemes used for the simulation of the plasma evolution in a tokamak.

\smallskip 

Succeeding to produce energy via thermonuclear fusion processes in a tokamak, is strongly dependent on the aptitude to confine the plasma in the core of the tokamak and at the same time on the ability to control the plasma heat-flux on the wall of the device. These two requirements (sine qua non) are incommoded by the turbulent plasma transport occurring in such high temperature environments. So, one of the most important research fields at the moment is the comprehension, the prediction as well as the control of the turbulent plasma flow in a tokamak, in particular in the edge region of such a device. What we mean with ``edge'' or ``peripheral'' region  of the tokamak is the region constituted firstly of the open magnetic field lines of the SOL (Scrape-off Layer), intercepting the wall (at the limiter), and secondly of the closed field line region, nearby the separatrix.

The study of peripheral tokamak plasmas is crucial for several reasons. First of all, this region imposes boundary conditions for the core plasma, and has thus to be treated with care. In particular, it is very important for the confinement properties of the reactor, to understand the turbulences occurring there. Secondly, it is the SOL-plasma which enters into contact with the wall of the reactor, such that controlling this peripheral plasma can be important for the protection of the reactor.

In this peripheral region, fluid models are often employed to describe the plasma evolution. This is due to the befalling smaller temperatures, which induce smaller particle mean-field paths such that closure relations, such as the Braginskii closures \cite{Braginskii}, are valid and permit to obtain macroscopic fluid models from more accurate kinetic ones. The TOKAM3X code \cite{Tamain1, Tamain2}, which is at the basis of the present work, is founded on such a fluid approach. The strong magnetic configuration has as an effect that the underlying conservation laws are strongly anisotropic. The plasma dynamics parallel to the magnetic field lines is very fast, whereas the dynamics in the perpendicular direction (with respect to the magnetic field) is more constraint. Multiple scales are thus present in the governing conservation laws, which render the mathematical as well as numerical study very challenging. Specific numerical algorithms have hence to be designed in order to capture accurately the physics of interest, without unnecessary small time and space steps. This is of primary importance, in order to reduce computational burden in such huge simulations.\\

The aim of the present paper is hence to contribute to the advance of the TOKAM3X code by proposing an efficient numerical scheme for a singularly-perturbed equation, which seems to be for the moment one of the troublesome points of this code. Indeed, the vorticity equation, which permits to compute the electric potential $\phi$, is strongly anisotropic, leading numerically to an ill-conditioned problem. Standard numerical techniques require refined meshes in order to account for the strong anisotropy and to get accurate results. This leads necessarily to excessive computational costs and multi-scale numerical schemes could be a welcome alternative. The {\it Asymptotic-Preserving} scheme we propose in this paper is based on a microscopic-macroscopic decomposition of the unknown function (electric potential $\phi$) separating in this manner the fast and rapid scales in the problem and permitting to transform the original singularly-perturbed problem into a regularly-perturbed one. This last one is then simply discretized by a standard finite-difference method. This procedure allows the choice of meshes independent on the perturbation parameter $\eta$, which describes the anisotropy, the numerical results remaining  uniformly accurate for all ranges of  $\eta \in [0,1]$.\\

Several types of AP-schemes have been introduced in literature, for various types of singularly-perturbed problems. To mention only some examples, we refer the interested reader to \cite{Crouseilles,Degond,filjin1,Jin,Jin_rev,Lemou,claudia_rev} and references therein. Briefly, {\it Asymptotic-Preserving} schemes are efficient procedures for solving singularly-perturbed problems. They consist in trying to mimic, at the discrete level, the asymptotic behaviour of the problem-solution as the perturbation parameter $\eta$ tends towards zero. This property renders the AP-scheme uniformly accurate, with respect to $\eta$.
The AP-scheme proposed in the present paper is based on the experience of the authors acquired with the study of highly anisotropic elliptic \cite{DDLNN,DDN,DLNN} or parabolic equations \cite{LNN,MN}. 

\bigskip 

The structure of this paper is the following: In Section \ref{SEC2} we shall present in detail the mathematical problem and the numerical difficulties one can encounter for its resolution. Section \ref{SEC3} presents and investigates mathematically an Asymptotic-Preserving reformulation for the underlying singularly-perturbed problem, reformulation which suites better for the limit of vanishing resistivity parameter, {\it i.e.} $\eta \rightarrow 0$. Section \ref{SEC4} deals then with the discretization of the AP-method we introduced above. And finally, in Section \ref{SEC5} we shall first validate the AP-scheme, and then apply it in the case of a real thermonuclear plasma experiment.

\section{The mathematical model} \label{SEC2} 

\begin{figure}
	\begin{center}
		\begin{tikzpicture}[text centered]
		
		\pgfmathsetmacro{\sx}{10}
		\pgfmathsetmacro{\sy}{5}
		\pgfmathsetmacro{\faxis}{0.3}
		
		\pgfmathsetmacro{\zo}{0*\sx}
		\pgfmathsetmacro{\za}{0.2*\sx}
		\pgfmathsetmacro{\zb}{0.8*\sx}
		\pgfmathsetmacro{\zu}{1*\sx}
		
		\pgfmathsetmacro{\ro}{0*\sy}
		\pgfmathsetmacro{\rl}{0.4*\sy}
		\pgfmathsetmacro{\rL}{1*\sy}
		
		\coordinate (O) at (\zo, \ro) ;
		\coordinate (Oy) at (\zo, \ro-\faxis*\ro+\faxis*\rl) ;
		\coordinate (Ox) at (\zo-\faxis*\zo+\faxis*\za, \ro) ;
		
		\coordinate (A) at (\za, \ro) ; 
		\coordinate (B) at (\zb, \ro) ;
		\coordinate (C) at (\zb, \rl) ;
		\coordinate (D) at (\zu, \rl) ; 
		\coordinate (E) at (\zu, \rL) ; 
		\coordinate (F) at (\zo, \rL) ;
		\coordinate (G) at (\zo, \rl) ;
		\coordinate (H) at (\za, \rl) ;
		
		\coordinate (GA) at (\zo, \ro) ;
		\coordinate (BD) at (\zu, \ro) ;
		
		\draw [<->, thick] (Oy) -- (O) -- (Ox) ;
		\node [below left] at (O) {$O$} ;
		
		\node [below, text width=1cm] at (Ox) {$z$\\(longitudinal)} ;
		
		\node [left] at (Oy) {(radial) $r$} ;
		
		\draw [ultra thick, fill=black!5!white] (A) -- (B) -- (C) -- (D) -- (E) -- (F) -- (G) -- (H) -- (A) ;
		\draw [ultra thick, color=red] (B) -- (C) (H) -- (A) ;
		\draw [ultra thick, color=blue] (D) -- (E) (F) -- (G) ;
		
		\node [below] at (A) {$z=a$} ;
		\node [below] at (B) {$z=b$} ;
		\node [below] at (BD) {$z=L_z$} ;
		
		\node [left] at (G) {$r=l$} ;
		\node [left] at (F) {$r=L_r$};
		
		\node [right] at (\za, 0.5*\ro+0.5*\rl) {$\Gamma_a$} ;
		\node [left]  at (\zb, 0.5*\ro+0.5*\rl) {$\Gamma_b$} ;
		
		\node [right] at (\zo, 0.5*\rl+0.5*\rL) {$\Gamma_0$} ;
		\node [left]  at (\zu, 0.5*\rl+0.5*\rL) {$\Gamma_{L_z}$} ;
		
		\node [above] at (0.5*\za+0.5*\zb, \ro) {$\Sigma_0$} ;
		\node [above] at (0.5*\zo+0.5*\za, \rl) {$\Sigma_l$} ;
		\node [above] at (0.5*\zb+0.5*\zu, \rl) {$\Sigma_l$} ;
		\node [below] at (0.5*\zo+0.5*\zu, \rL) {$\Sigma_{L_r}$} ;
		
		\node at (0.5*\zo+0.5*\zu, 0.5*\ro+0.5*\rL) {$\Omega$} ;
		
		\node at (0.5*\za, 0.5*\ro+0.5*\rl) {$limiter$} ;
		\node at (0.5*\zb+0.5*\zu, 0.5*\ro+0.5*\rl) {$limiter$} ;
		
		\draw [thick, dashed] (G) -- (GA) -- (A) ;
		\draw [thick, dashed] (B) -- (BD) -- (D) ;
		
		\end{tikzpicture}
	\end{center}
	\caption{The 2D domain $\Omega$, representing the SOL plasma region. $z$ is the longitudinal coordinate and $r$ is the radial coordinate.\label{fig:domain}}
\end{figure}
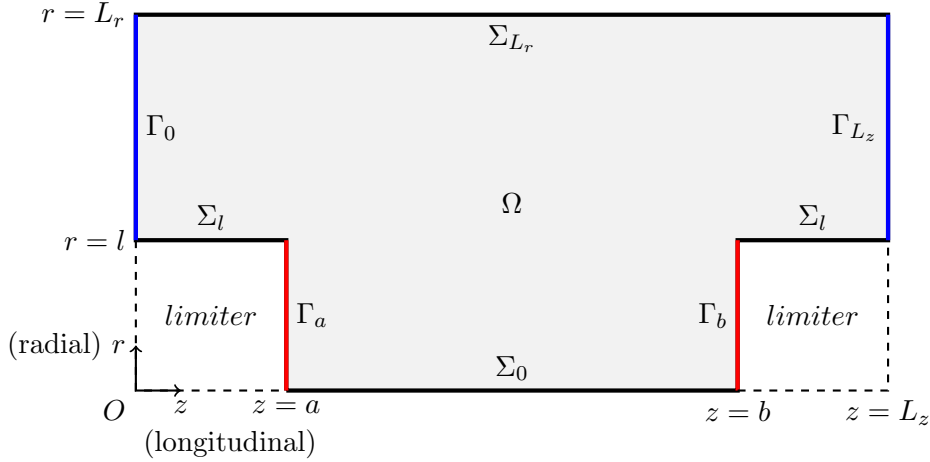

Let us now present the singularly-perturbed, non-linear problem, describing the evolution of the electric potential $\phi$ in the peripheral tokamak region, represented by the domain $\Omega \subset \RR^2$, which is sketched in Fig.~\ref{fig:domain}. This equation reads
\begin{equation} \label{EQev}
-\partial_t \partial^2_r \phi -\frac{1}{\eta}\partial^2_z \phi
+\nu\partial^4_r \phi  = {\mathcal S} - {1 \over \eta} \partial_z {\mathcal F}\,, \quad t\ge0\,, \quad (r,z) \in \Omega\,,
\end{equation}
and is completed with the initial condition
\begin{equation} \label{IC}
\partial_r \phi(0,r,z)=\partial_r \phi_{in}(r,z)\,, \quad (r,z) \in \Omega\,,
\end{equation}
for some given function $\phi_{in}$. The imposed boundary conditions are the no-slip boundary conditions on $\Sigma \coloneqq \Sigma_0 \cup\Sigma_l \cup \Sigma_{L_r}=\{ (r,z) \in \partial \Omega \,\, / \,\, r=0, \, r=l \,\, \textrm{or}\,\, r=L_r\}$
\begin{equation} \label{BC1}
\partial_r \phi (t,r,z) = \partial^3_r \phi(t,r,z)= 0\,, \quad t\ge 0\,, \quad (r,z)
\in \Sigma \,,
\end{equation}
periodic boundary conditions on $\Gamma_0\cup \Gamma_{L_z}=\{ (r,z) \in \partial \Omega \,\, / \,\, z=0 \,\, \textrm{or}\,\, z=L_z\}$, that means in the core of the plasma, and the nonlinear sheath boundary conditions
on the limiters $\Gamma_a \cup \Gamma_b=\{ (r,z) \in \partial \Omega \,\, / \,\, z=a \,\, \textrm{or}\,\, z=b\}$
\begin{equation} \label{BCNL}
\left\{
\begin{array}{l}
\displaystyle \partial_z \phi(t,r,a)= \eta(1-e^{\Lambda-\phi(t,r,a)}) + {\mathcal F} (t,r,a)\,, \quad
t\ge 0\,, \quad (r,z) \in \Gamma_a\,, \\[3mm]
\displaystyle \partial_z \phi(t,r,b)=-\eta(1-e^{\Lambda-\phi(t,r,b)}) +{\mathcal F}(t,r,b)\,, \quad
t\ge 0\,, \quad (r,z) \in \Gamma_b\,.
\end{array}
\right.
\end{equation}
The source term is composed of a stiff and a non-stiff part, denoted respectively by $\partial_z{\mathcal F}$ and  ${\mathcal S}$, whereas $\eta>0$, $\nu>0$ and $\Lambda>0$ are some given constants. The parameter $\Lambda$ is the sheath floating potential and $\eta$ represents the parallel resistivity of the plasma and is supposed to be very small $0 < \eta \ll 1$, resulting in a singularly-perturbed problem, which is very challenging to solve numerically. The domain $\Omega$ covers both, closed flux surfaces (periodic region) and open flux surfaces, touching the limiter (nonlinear boundary conditions), describing thus the so-called SOL region (Scrape-off Layer) of a tokamak plasma.\\

Our model problem \eqref{EQev}--\eqref{BCNL} is extracted from the TOKAM3X model \cite{Tamain1,Tamain2} which describes the dynamics of a magnetically confined tokamak edge-plasma via a fluid approach. 
At the basis of the TOKAM3X code are the balance equations for the electrons and ions, coupled to the Poisson equation for the electrostatic potential. Under some suitable assumptions, such as for example the quasi-neutrality condition, the low mass ratio $m_e/m_i$ assumption, the drift-approximation, and so on, one obtains a fluid model, based on the particle balance equation, the parallel momentum equation, the charge balance ($\nabla \cdot j =0$) and the parallel Ohm's law, describing the evolution of respectively the electron density $N$, the parallel ion momentum $\Gamma$, the plasma potential $\phi$ and the parallel current $j_{||}$, parallel with respect to the imposed strong magnetic field. Introducing the vorticity quantity $W$, the set of equations at the foundation of the TOKAM3X code, read
\be \label{TOK}
\left\{
\begin{array}{l}
\ds \partial_t N  + \nabla \cdot (N\, {\mathbf u_e}) - \nabla \cdot ( D_N \, \nabla_\perp N) = S_N\,,\\[3mm]
\ds \partial_t \Gamma + \nabla \cdot (\Gamma \, {\mathbf u_i}) - \nabla \cdot ( D_\Gamma \, \nabla_\perp \Gamma) = -\nabla_{||} P\,,\\[3mm]
\ds \partial_t W+ \nabla \cdot (W\, {\mathbf u_i}) - \nabla \cdot ( D_W \, \nabla_\perp W) = \nabla \cdot [N( {\mathbf u^i_{\nabla B}} - {\mathbf u^e_{\nabla B}})] + j_{||}\ {\mathbf b}\,,
\end{array}
\right.
\ee
with the vorticity $W$ and the parallel current $j_{||}$ defined as 
$$
W \coloneqq \nabla \cdot \left[{ 1 \over |{\mathbf B}|^2} \, ( \nabla_\perp \phi + { 1 \over N}\, \nabla_\perp N)\right]\,, \qquad 
j_{||} \coloneqq - { 1 \over \eta_{||}} \, \nabla_{||} \phi + { 1 \over \eta_{||}\, N} \, \nabla_{||} N\,.
$$
In this system ${\mathbf B}$ is the magnetic field, with direction ${\mathbf b} \coloneqq {{\mathbf B} \over |{\mathbf B}|}$, $S_N$ is a source term, $P$ is the static pressure defined as $P \coloneqq N\, (T_i+T_e)$ and $\eta_{||}$ is the normalized parallel collisional resistivity of the plasma. Furthermore ${\mathbf u_{i,e}}={ u^{i,e}}_{||}\,{\mathbf b} + {\mathbf u^{i,e}}_{\perp} $ denote the particle macroscopic velocities, with the perpendicular parts ${\mathbf u^{i,e}}_{\perp}={\mathbf u}_{\mathbf E}+{\mathbf u^{i,e}_{\nabla B}}$ characterized in terms of the electric and curvature drift velocities, given by
$$
{\mathbf u}_{\mathbf E}:= {{\mathbf E} \times {\mathbf B} \over |{\mathbf B}|^2}\, \quad  \quad {\mathbf u^{i,e}_{\nabla B}}:= \pm { 2\, T_{i,e} \over e |{\mathbf B}|}\, {({\mathbf B} \times \nabla  {\mathbf B}) \over |{\mathbf B}|^2}\,,
$$
and where $T_{i,e}$ are the respective particle temperatures, considered in the present case as constant. The diffusion coefficients $D_{N,\Gamma,W}$ account for the collisional transport and/or diffusive transport (neoclassical and anomalous) and model the turbulences at small scales. System \eqref{TOK} allows to study the isothermal, electrostatic, 3D turbulences arising in the SOL. For its detailed derivation from the underlying conservation laws, we refer the interested reader to the references \cite{Tamain1,Tamain2}. The numerical simulations as well as a detailed analysis of the obtained results are also presented there. The authors remark that there are still some numerical difficulties to be further inquired in the vorticity equation, and this due to the strong anisotropy of the problem. In order to address these numerical difficulties, we decided to extract in this paper the vorticity equation from the whole system \eqref{TOK} and simplify it, keeping only the terms which cause complications. A simplified version of this potential equation is given by our problem \eqref{EQev}--\eqref{BCNL} (see previous work \cite{NNGS} for its derivation from \eqref{TOK}).\\

The rigorous mathematical study, meaning the study of the well-posedness of this problem $\forall \eta >0$ (existence, uniqueness and stability of a weak solution), has been considered in \cite{NNGS}. We observed in that work that the primary mathematical difficulties (for fixed $\eta >0$) arise on one hand from the degeneracy in time of the problem (regularization) and on the other hand from the non-linear boundary conditions. These difficulties will naturally also occur in the numerical treatment of this problem. \\

The main goal of the present paper is to propose a numerical follow-up of the latter more mathematical paper, in the aim to design an efficient numerical scheme for the resolution of problem \eqref{EQev}--\eqref{BCNL}. Typical computational challenges within this task come from:\\
\begin{itemize}
\item the high anisotropy of the problem, described by the small resistivity parameter $\eta \ll 1$, and introduced into the problem by the strong magnetic field, which confines the plasma;
\item the non-linearity of the boundary conditions, describing the plasma-wall interactions (Bohm-conditions).
\end{itemize}

Concerning the first point, the anisotropy, we have to deal with a typical singularly-perturbed problem when $\eta \rightarrow 0$. Such kind of problems are particularly difficult to treat numerically, as the equations change type as the perturbation parameter $\eta$ tends towards zero, leading (usually) to ill-posed problems in the limit. In the present case, the so-called ``reduced-model'' has the form
\be \label{RR}
(R)\,\,\, \left\{
\begin{array}{l}
\ds -\partial^2_z \phi  = -  \partial_z {\mathcal F}\,, \quad t\ge0\,, \quad (r,z) \in \Omega\,,\\[3mm]
\displaystyle \partial_z \phi(t,r,a)=  {\mathcal F} (t,r,a)\,, \quad
t\ge 0\,, \quad (r,z) \in \Gamma_a\,, \\[3mm]
\displaystyle \partial_z \phi(t,r,b)={\mathcal F}(t,r,b)\,, \quad
t\ge 0\,, \quad (r,z) \in \Gamma_b\,,
\end{array}
\right.
\ee
associated with the other boundary conditions on $\Sigma$ and $\Gamma_0 \cup \Gamma_{L_z}$ as well as the initial conditions. One can now remark that this reduced model is ill-posed, as it admits either no solution (if the initial condition is not well-prepared) or an infinite amount of solutions, as one can add to any solution another $r$-dependent function, satisfying the boundary conditions on $\Sigma$ and the initial condition. On the discrete level, the $(R)$-problem will always have infinitely many solutions, as one steps over the initial condition. This distinction between well-prepared and not well-prepared initial condition is related to the creation of a boundary layer near $t=0$.\\

At the discrete level, all these complications will be translated into the fact that the linear system to be solved will become singular, as $\eta \rightarrow 0$, or equally, becomes ill-conditioned as $\eta \ll 1$, leading hence to erroneous results. For not too small, fixed $\eta$-values, a preconditioner could help, however in a general case, where the perturbation parameter $\eta$ varies within the simulation domain $\Omega$ and takes various orders of magnitude, the preconditioner will no longer follow and new techniques have to be employed to rescue the user.\\

The occurrence of all these difficulties (theoretical as well as numerical) is strongly related to the multi-scale character of our problem. Indeed, for small $\eta \ll 1$ the problem is evolving more rapidly in the $z$-direction, which represents the direction of the strong magnetic field, than in the perpendicular $r$-direction. Different space- and time-scales are hence introduced in the problem by the perturbation parameter $\eta$, and in order to be accurate, a standard numerical scheme has to take into account for all these small scales, by imposing very restrictive grid-conditions as for example meshes of order $\eta$. This can become rapidly too costly.\\

The aim of the next sections will be hence to introduce a multi-scale numerical scheme, based on the study of the asymptotic behaviour of the solution $\phi^\eta$ as $\eta$ goes to zero. A decomposition of $\phi^\eta$ into a macroscopic part and a microscopic one separates somehow the different dynamics in the problem.
This procedure permits then the use of judicious mesh-sizes and time-steps, adapted to the physical phenomenon one wants to study and not to the perturbation parameter and permits to treat with no huge computational costs even the limiting case $\eta \equiv 0$.

\section{An asymptotic-preserving reformulation for the evolution problem}\label{SEC3}
This section is devoted to a reformulation of the original singularly-perturbed electric potential problem \eqref{EQev}--\eqref{BCNL}, denoted in the sequel by $(SP)_\eta$, into a problem which behaves better in the limit $\eta \rightarrow 0$.\\

The essence of our numerical method for the resolution of \eqref{EQev}--\eqref{BCNL} is based on the following micro-macro decomposition of the unknown $\phi^\eta$ 
\be \label{decomp}
\phi^\eta=p^\eta + \eta q^\eta\,, \quad \quad \forall \eta >0\,,
\ee
where the macroscopic function $p^\eta$ is chosen to be solution of the ``dominant'' problem
\be \label{PB_p}
\left\{
\begin{array}{l}
\ds -\partial_{z}^2 p^\eta = - \partial_z {\mathcal F}\,, \quad t\ge0\,, \quad (r,z) \in \Omega\,,\\[3mm]
\ds  \partial_{z} p^\eta = {\mathcal F} \quad \textrm{on} \quad \Gamma_a \cup \Gamma_b\,,
\end{array}
\right.
\ee
associated with the other homogeneous or periodic conditions on the remaining boundaries. For the uniqueness of the decomposition, we shall further fix $q^\eta$ (or equivalently $p^\eta$) on an interface, denoted here $\Gamma_q$, as follows
\be  \label{PB_q}
q^\eta_{| \Gamma_{q}} \equiv q_\star \quad \textrm{on} \quad \Gamma_{q} \coloneqq \{ (r,z)\in \Omega\,\, / \,\, z={a+b \over 2}\,, \,\, r \in [0,L] \}\,\,\, \Rightarrow\,\,\, p^\eta_{| \Gamma_{q}} = \phi^\eta_{| \Gamma_{q}}-\eta q_\star\,.
\ee
Indeed, the decomposition \eqref{decomp} is now unique for given $\phi^\eta$ and fixed $\eta >0$, precisely due to the fact that we impose $q^\eta$ on $\Gamma_q$, fixing thus also $p^\eta$ on this interface. Problem \eqref{PB_p} becomes thus a well-posed elliptic problem with unique solution $p^\eta$ for all $\eta >0$.\\

With this decomposition, the problem $(SP)_\eta$ transforms now into the completely equivalent system
\begin{equation} \label{MM}
(AP)_\eta\,\,\, \left\{
\begin{array}{l}
\ds -\partial_t \partial^2_r \phi^\eta -\partial^2_z q^\eta
+\nu\partial^4_r \phi^\eta  = {\mathcal S}\,, \quad t\ge0\,, \quad (r,z) \in \Omega\,,\\[3mm]
\ds - \partial^2_z \phi^\eta = - \eta \partial^2_z q^\eta -\partial_z {\mathcal F}\,, \quad t\ge0\,, \quad (r,z) \in \Omega\,,
\end{array}
\right.
\end{equation}
associated with the usual initial condition for $\phi^\eta$, the usual homogeneous resp. periodic boundary conditions on $\Sigma$ resp. $\Gamma_0 \cup \Gamma_{L_z}$ and the following  boundary conditions on $\Gamma_a \cup \Gamma_b$ for the unknowns $(\phi^\eta,q^\eta)$ 
\be \label{AP_BC}
\left\{
\begin{array}{l}
\ds q^\eta_{| \Gamma_{q}} \equiv q_\star\\[3mm]
\ds \partial_z q^\eta_{| \Gamma_{a}}= (1-e^{\Lambda-\phi^\eta(t,r,a)})\\[3mm]
\ds \partial_z q^\eta_{| \Gamma_{b}}= -(1-e^{\Lambda-\phi^\eta(t,r,b)})\\[3mm]
\ds \partial_z \phi^\eta_{| \Gamma_{a}}= \eta (1-e^{\Lambda-\phi^\eta(t,r,a)})+ {\mathcal F}(t,r,a)\\[3mm]
\ds \partial_z \phi^\eta_{| \Gamma_{b}}= -\eta (1-e^{\Lambda-\phi^\eta(t,r,b)})+ {\mathcal F}(t,r,b)\,.
\end{array}
\right.
\ee
Note that no boundary condition for $q^\eta$ is needed on $\Sigma$, as no $r$-derivatives of $q^\eta$ occur in the system. We shall call in the following this problem the Asymptotic-Preserving reformulation of our Singularly-Perturbed problem \eqref{EQev}--\eqref{BCNL}, denoted simply by $(AP)_\eta$.\\

The equivalence of both problems, $(SP)_\eta$ and $(AP)_\eta$, for any $\eta >0$, is due to the uniqueness of the solution of \eqref{PB_p} when imposing $p^\eta_{|\Gamma_q}= \phi^\eta_{|\Gamma_q} - \eta \, q_\star$. This equivalence together with the mathematical existence and uniqueness studies of the problem $(SP)_\eta$, considered in \cite{NNGS}, permits to show that $(AP)_\eta$ is well-posed for each $\eta >0$. The essential difference between these two reformulations is perceived only in the limit $\eta \rightarrow 0$. Indeed, $(SP)_\eta$ becomes singular, as explained in Section \ref{SEC2}. Let us yet formally investigate what happens with the micro-macro reformulation $(AP)_\eta$, when $\eta$ goes to zero. Setting formally $\eta \equiv 0$ one gets the system
\begin{equation} \label{MM_0}
(AP)_0\,\,\, \left\{
\begin{array}{l}
\ds -\partial_t \partial^2_r \phi^{0} -\partial^2_z q^{0}
+\nu\partial^4_r \phi^{0}  = {\mathcal S}\,, \quad t\ge0\,, \quad (r,z) \in \Omega\,,\\[3mm]
\ds - \partial^2_z \phi^{0} = -\partial_z {\mathcal F}\,, \quad t\ge0\,, \quad (r,z) \in \Omega\,,
\end{array}
\right.
\end{equation}
associated with the following  boundary conditions for the unknowns $(\phi^{0}, q^{0})$
$$
\left\{
\begin{array}{l}
\ds q^{0}_{| \Gamma_{q}} \equiv q_\star\\[3mm]
\ds \partial_z q^{0}_{| \Gamma_{a}}= (1-e^{\Lambda-\phi^{0}(t,r,a)})\\[3mm]
\ds \partial_z q^{0}_{| \Gamma_{b}}= -(1-e^{\Lambda-\phi^{0}(t,r,b)})\\[3mm]
\ds \partial_z \phi^{0}_{| \Gamma_{a}}=  {\mathcal F}(t,r,a)\\[3mm]
\ds \partial_z \phi^{0}_{| \Gamma_{b}}=  {\mathcal F}(t,r,b)\,,
\end{array}
\right.
$$
which is a type of saddle-point problem. The unknown $q^{0}$ can be seen here as a Lagrangian multiplier, associated to the constraint $- \partial^2_z \phi^{0} = -\partial_z {\mathcal F}$. The rigorous well-posedness of this limit problem is not the aim of the present paper, which is much more numerical, can however be a nice extension of this work, involving saddle-point theory.\\
Hence, the main benefit of the AP-reformulation $(AP)_\eta$ is that in the limit $\eta \rightarrow 0$ one gets a well-posed problem, such that we have no more to face numerical singularities, when solving the reformulation for small $\eta$-values.  This big advantage shall be extensively put into light with the simulations performed in Section \ref{SEC5}.\\

Before proceeding to the numerical treatment, let us mention here some words about the essence of the micro-macro decomposition \eqref{decomp}, which is at the basis of our AP-reformulation. The idea behind was to eliminate the dominant, stiff operator, and this has been done by introducing a sort of separation of scales. The function $p^\eta$ solves the dominant operator, whereas $q^\eta$ incorporates the microscopic information. In the limit $\eta \rightarrow 0$ the microscopic part $q^0$ is still present in the homogenized limit model $(AP)_0$, and it is this part which renders the problem well-posed. It permits to recover the microscopic information, which was lost in the reduced model \eqref{RR}.

\section{The numerical discretization via finite differences}\label{SEC4}

In contrast to previous papers on highly anisotropic elliptic or parabolic problems \cite{DDLNN, DDN, DLNN, LNN, MN}, we made here the choice to solve the model equations outlined in Eqs.~\eqref{EQev}--\eqref{BCNL} and \eqref{MM}--\eqref{AP_BC} by means of a numerical approach based on finite difference approximations, instead of relying on the finite element method. The reason for this choice is the fact that we would like to provide the team developing the TOKAM3X code, based on a discretization of the balance equations via the finite volume method, with a technique directly applicable in that context in a way as straightforward as possible.

\smallskip 

For the investigation of the consistency and accuracy of the numerical scheme  proposed here, a Cartesian uniform mesh with constant mesh size along both longitudinal and radial directions ($\Delta r \equiv \Delta z = const$) has been adopted (see Section \ref{SEC52}).
This constant uniform mesh may however be abandoned to introduce mesh refinement near the borders of the domain, or where steep gradients of the solution are to be expected. Such a non-uniform Cartesian mesh has been adopted in the second part of our investigation, when the analysis of a problem setting inspired by a real physical plasma application is proposed. This will be presented in Section \ref{SEC53}.

\smallskip 

The semi-discretization in space is not the most critical part in the construction of an AP scheme, though. In fact, special care must be paid to the time-discretization, in particular when decisions have to be taken concerning which terms to take implicitly and which ones explicitly. In order not to destroy the desirable properties of our AP formulation, the time discretization is here based on the implicit Euler scheme. The interval $[0,\,T]$ is discretized in uniform steps $\Delta t$, the solution being evaluated at the time instants $t_0, t_1, \ldots t_{N_t}$ defined as 
$$
	t_k \coloneqq k\, \Delta t, \qquad\qquad k=0, \ldots, N_t, \qquad\qquad \Delta t \coloneqq T/{N_t}. 
$$

\medskip 
Let us present now the discretizations of both formulations, the {\APeta }-scheme formulated in Eqs.~\eqref{MM}--\eqref{AP_BC}  respectively the {\SPeta }-scheme formulated in Eqs.~\eqref{EQev}--\eqref{BCNL}. In the following Section \ref{SEC5}, we shall then compare the performances of these two formulations.

\subsection{Semi-discretization in time of {\SPeta}.}\label{SEC41}

Discretizing Eq.~\eqref{EQev} in time by means of the implicit Euler scheme, yields
\begin{equation}
	-\partial_r^2 \left( \frac{\phi^{n+1} - \phi^{n}}{\Delta t} \right) - \frac{1}{\eta} \partial_z^2 \phi^{n+1} + \nu\, \partial_r^4 \phi^{n+1} = {\mathcal S}^{n+1} - {1 \over \eta} \partial_z {\mathcal F}^{n+1},
\end{equation}
and hence for $n=0, \ldots, N_t-1$,
\begin{equation} \label{eq:SP-discr-0}
	-\partial_r^2 \phi^{n+1} - \frac{\Delta t}{\eta} \partial_z^2 \phi^{n+1} + \nu \Delta t\, \partial_r^4 \phi^{n+1} = \Delta t\, {\mathcal S}^{n+1} - {\Delta t \over \eta} \partial_z {\mathcal F}^{n+1}- \partial_r^2 \phi^{n},
\end{equation}
where $\phi^k$, $\mathcal{S}^k$ and $\mathcal{F}^k$ denote the solution and the source terms $\mathcal{S}$ resp. $\mathcal{F}$, evaluated at the $k$-th time step, and $\phi^0$ is given by the initial condition $\phi_{in}$ \eqref{IC}. 

This equation is associated with the following nonlinear boundary conditions:
\begin{equation} \label{eq:SP-discr-1}
\left\{
\begin{split}
	& \partial_r \phi^{n+1} = \partial^3_r \phi^{n+1} = 0 \quad \text{on} \quad \Sigma_0 \cup\Sigma_l \cup \Sigma_{L_r};\\
	& \text{periodic boundary conditions for $\phi^{n+1}$ on $\Gamma_0\cup \Gamma_{L_z}$};\\
	& \partial_z \phi^{n+1}_a = \eta \left( 1-e^{\Lambda-\phi^{n+1}_a}\right) + {\mathcal F}_a^{n+1} \quad \text{on} \quad\Gamma_a; \\
	& \partial_z \phi^{n+1}_b =  - \eta \left( 1-e^{\Lambda-\phi^{n+1}_b}\right) + {\mathcal F}_b^{n+1} \quad \text{on} \quad \Gamma_b;
\end{split}
\right.
\end{equation}
%
where we used the notation $\phi^{n+1}_a \equiv \phi^{n+1}\left(t,r,z=a\right)$, $\phi^{n+1}_b \equiv \phi^{n+1}\left(t,r,z=b\right)$ and similarly for ${\mathcal F}^{n+1}$.

\subsection{Semi-discretization in time of {\APeta}.}\label{SEC42}

Discretizing Eq.~\eqref{MM} in time by means of the implicit Euler scheme yields for $n = 0, 1, \ldots, N_t-1$
\begin{equation} \label{eq:AP1-discr-0}
\left\{
\begin{split}
	&-\partial_r^2 \phi^{n+1} - \Delta t\, \partial_z^2 q^{n+1} + \nu \Delta t\, \partial_r^4 \phi^{n+1} =  \Delta t\, {\mathcal S}^{n+1} - \partial_r^2 \phi^{n},\\[3mm]
&-\partial_z^2 \phi^{n+1} + \eta \partial_z^2 q^{n+1} =  -\partial_z {\mathcal F}^{n+1},
\end{split}
\right.
\end{equation}
where $q^{n+1}$ denotes the microscopic function $q$ at the $(n+1)$-th time step. Eq.~\eqref{eq:AP1-discr-0} is associated with the following nonlinear boundary conditions:
\begin{equation} \label{eq:AP1-discr-1}
\left\{
\begin{split}
	& \partial_r \phi^{n+1} = \partial^3_r \phi^{n+1} = 0 \quad \text{on} \quad \Sigma_0 \cup\Sigma_l \cup \Sigma_{L_r}; \\
	& \text{periodic boundary conditions for both $\phi^{n+1}$ and $q^{n+1}$ on $\Gamma_0\cup \Gamma_{L_z}$}; \\
	& \partial_z \phi^{n+1}_a = \eta \left( 1-e^{\Lambda-\phi^{n+1}_a}\right) + {\mathcal F}_a^{n+1} \quad \text{on} \quad \Gamma_a; \\
	& \partial_z \phi^{n+1}_b =  - \eta \left( 1-e^{\Lambda-\phi^{n+1}_b}\right) + {\mathcal F}_b^{n+1} \quad \text{on} \quad \Gamma_b; \\
	& \partial_z q^{n+1}_a = \left(1-e^{\Lambda-\phi^{n+1}_a}\right) \quad \text{on} \quad \Gamma_a;\\
	& \partial_z q^{n+1}_b =  - \left( 1-e^{\Lambda-\phi^{n+1}_b}\right) \quad \text{on} \quad \Gamma_b;\\
	& q^{n+1} \equiv q_{\star} \quad \text{on} \quad\Gamma_q,
\end{split}
\right.
\end{equation}
where $q_{\star}$ is an arbitrary constant.

\subsection{Linearization of the boundary conditions.}\label{SEC411}
The treatment of the nonlinear boundary conditions is the same for both formulations {\SPeta } or {\APeta}, and is based on a linearization obtained via a Taylor expansion. The following approximation has been adopted:
\begin{equation}
e^{\Lambda - \phi_\alpha^{n+1}} = %
e^{\Lambda - \phi_\alpha^n}\, e^{\phi_\alpha^n- \phi_\alpha^{n+1} } \simeq %
e^{\Lambda - \phi_\alpha^n} \left( 1 +\phi_\alpha^n- \phi_\alpha^{n+1}  \right) \qquad \left( \alpha = a, b\right),
\end{equation}
leading to the following Robin boundary condition, substituting the boundary conditions on $\Gamma_a$ and $\Gamma_b$ for the {\SPeta } scheme (see Section~\ref{SEC41}),
\begin{equation} \label{eq:BC-SP}
\left\{
\begin{split}
	& \partial_z \phi^{n+1}_a  - \eta\, e^{\Lambda - \phi_a^n} \phi_a^{n+1} = \eta \left[ 1 - e^{\Lambda - \phi_a^n} \left( 1 + \phi_a^n \right) \right]+ {\mathcal F}_a^{n+1} \quad \text{on} \quad \Gamma_a, \\
	& \partial_z \phi^{n+1}_b+ \eta \, e^{\Lambda - \phi_b^n} \phi_b^{n+1}  =  - \eta \left[ 1 - e^{\Lambda - \phi_b^n} \left( 1 + \phi_b^n \right) \right]+ {\mathcal F}_b^{n+1} \quad \text{on} \quad \Gamma_b,
\end{split} 
\right.
\end{equation}
and the following conditions substituting the boundary conditions on $\Gamma_a$ and $\Gamma_b$ for the {\APeta } scheme (see Section~\ref{SEC42}):
\begin{equation} \label{eq:BC-AP}
\left\{
\begin{split}
	& \partial_z \phi^{n+1}_a  - \eta\, e^{\Lambda - \phi_a^n} \phi_a^{n+1} = \eta \left[ 1 - e^{\Lambda - \phi_a^n} \left( 1 + \phi_a^n \right) \right]+ {\mathcal F}_a^{n+1} \quad \text{on} \quad \Gamma_a,\\
	& \partial_z \phi^{n+1}_b+ \eta \, e^{\Lambda - \phi_b^n} \phi_b^{n+1}  =  - \eta \left[ 1 - e^{\Lambda - \phi_b^n} \left( 1 + \phi_b^n \right) \right]+ {\mathcal F}_b^{n+1} \quad \text{on} \quad \Gamma_b,\\
	& \partial_z q^{n+1}_a  - \, e^{\Lambda - \phi_a^n} \phi_a^{n+1} =\left[ 1 - e^{\Lambda - \phi_a^n} \left( 1 + \phi_a^n \right) \right] \quad \text{on} \quad \Gamma_a,\\
	& \partial_z q^{n+1}_b+ \, e^{\Lambda - \phi_b^n} \phi_b^{n+1}  =  - \left[ 1 - e^{\Lambda - \phi_b^n} \left( 1 + \phi_b^n \right) \right] \quad \text{on} \quad \Gamma_b.
\end{split}
\right.
\end{equation}

\subsection{Semi-discretization in space. \label{SEC43}}

The computational domain $\Omega$, sketched in Fig.~\ref{fig:domain}, has been discretized by means of a structured Cartesian grid.
We shall denote by $M_z$ the maximum number of grid nodes in the $z$ (longitudinal) direction (the node located on $z=L_z$ is not included in the computational domain -- and hence also in $M_z$ -- because of the periodic boundary conditions on $\Gamma_0$ and $\Gamma_{L_z}$), and by $M_r$ the maximum number of grid nodes in the $r$ (radial) direction. Since the domain is not a square, the total number of grid points, denoted by $M$, is in general $M < M_z M_r$. 

In the case of the {\SPeta } scheme, Eq.~\eqref{eq:SP-discr-0} is discretized in the $M$ grid points of the computational domain, leading to a system of algebraic equations $\mathbf{A} \mathbf{x} = \mathbf{b}$ with a number $M_U$ of unknowns matching the number of grid nodes, i.e.  $M_U=M$, where the components of the vector $\mathbf{x}$ represent the grid values of the function $\phi$.

In the case of the {\APeta } scheme, Eq.~\eqref{eq:AP1-discr-0}$_1$ is discretized on the $M$ grid points of the computational domain, and Eq.~\eqref{eq:AP1-discr-0}$_2$ is discretized on the $M-M_r$ grid nodes of the computational domain except the interface $\Gamma_q$ (due to the fact that the value of the function $q$ is prescribed on $\Gamma_q$, as explained in Section \ref{SEC3}), leading to $M_U=2M-M_r$ unknowns. The components of the vector $\mathbf{x}$ represent in this case the $M$ grid values of the function $\phi$ and the $M-M_r$ grid values of the microscopic function $q$. 

 
Standard finite difference approximations have been used to discretize the second and fourth order derivatives appearing in Eq.~\eqref{eq:SP-discr-0} and Eq.~\eqref{eq:AP1-discr-0}. Denoting with $u_{i,j}$ the value of the unknown variable $u$ ($u=\phi, q$) in the node located in the $i$-th column and $j$-th row of the Cartesian grid, whose $z$ and $r$ coordinates are, respectively, $z_i$ and $r_j$, we have 
\begin{equation*}
\begin{split}
	\left(\partial^2_z u\right)_{i,j} =& a_{i}^{-} u_{i-1,j} - \left(a_{i}^{-} + a_{i}^{+}\right) u_{i,j} + a_{i}^{+} u_{i+1,j}, \\
	\left(\partial^2_r u\right)_{i,j} =& b_{j}^{-} u_{i,j-1} - \left(b_{j}^{-} + b_{j}^{+}\right) u_{i,j} + b_{j}^{+} u_{i,j+1}, \\
	\left(\partial^4_r u\right)_{i,j} =& \left( b_{j-1}^{-} b_{j}^{-} \right) u_{i,j-2} - b_{j}^{-} \left( b_{j-1}^{-} + b_{j-1}^{+} + b_{j}^{-} + b_{j}^{+} \right) u_{i,j-1} +	\\
	&\quad +\left( b_{j-1}^{+} b_{j}^{-} + \left( b_{j}^{-} + b_{j}^{+} \right)^2 + b_{j}^{+} b_{j+1}^{-} \right) u_{i,j} +\\
	&\quad - b_{j}^{+} \left( b_{j}^{-} + b_{j}^{+} + b_{j+1}^{-} + b_{j+1}^{+} \right) u_{i,j+1} + \left( b_{j}^{+} b_{j+1}^{+} \right) u_{i,j+2},
\end{split}
\end{equation*}
with
\begin{gather}
	a_{i}^{-} = \frac{2}{\Delta z_i^{-} \left(\Delta z_i^{-} + \Delta z_i^{+}\right)}, \qquad 
	a_{i}^{+} = \frac{2}{\Delta z_i^{+} \left(\Delta z_i^{-} + \Delta z_i^{+}\right)} \qquad i=1, \ldots M_z, \\
	b_{j}^{-} = \frac{2}{\Delta r_j^{-} \left(\Delta r_j^{-} + \Delta r_j^{+}\right)}, \qquad 
	b_{j}^{+} = \frac{2}{\Delta r_j^{+} \left(\Delta r_j^{-} + \Delta r_j^{+}\right)} \qquad j=1, \ldots M_r,
\end{gather}
and
\begin{equation}
	\Delta z_i^{-} = z_{i} - z_{i-1}, \quad \Delta z_i^{+} = z_{i+1} - z_{i}, \quad
	\Delta r_j^{-} = r_{j} - r_{j-1}, \quad \Delta r_j^{+} = r_{j+1} - r_{j}.
\end{equation}
The boundary conditions given in Eq.~\eqref{eq:BC-SP} (for the {\SPeta } scheme) or Eq.~\eqref{eq:BC-AP} (for the {\APeta } scheme) are then properly used to eliminate the ghost-nodes.\\

\begin{figure}
	\includegraphics[scale=0.6]{\figdir/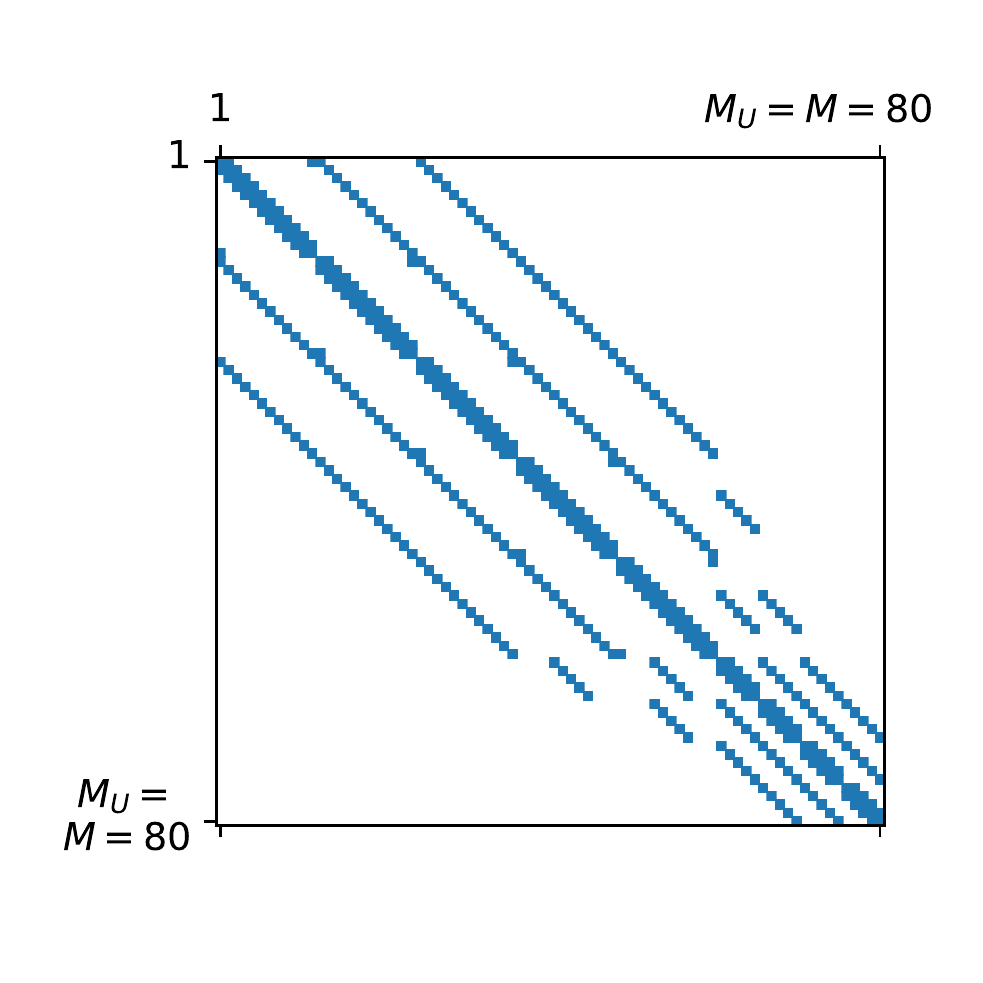} \ \ 
	\includegraphics[scale=0.6]{\figdir/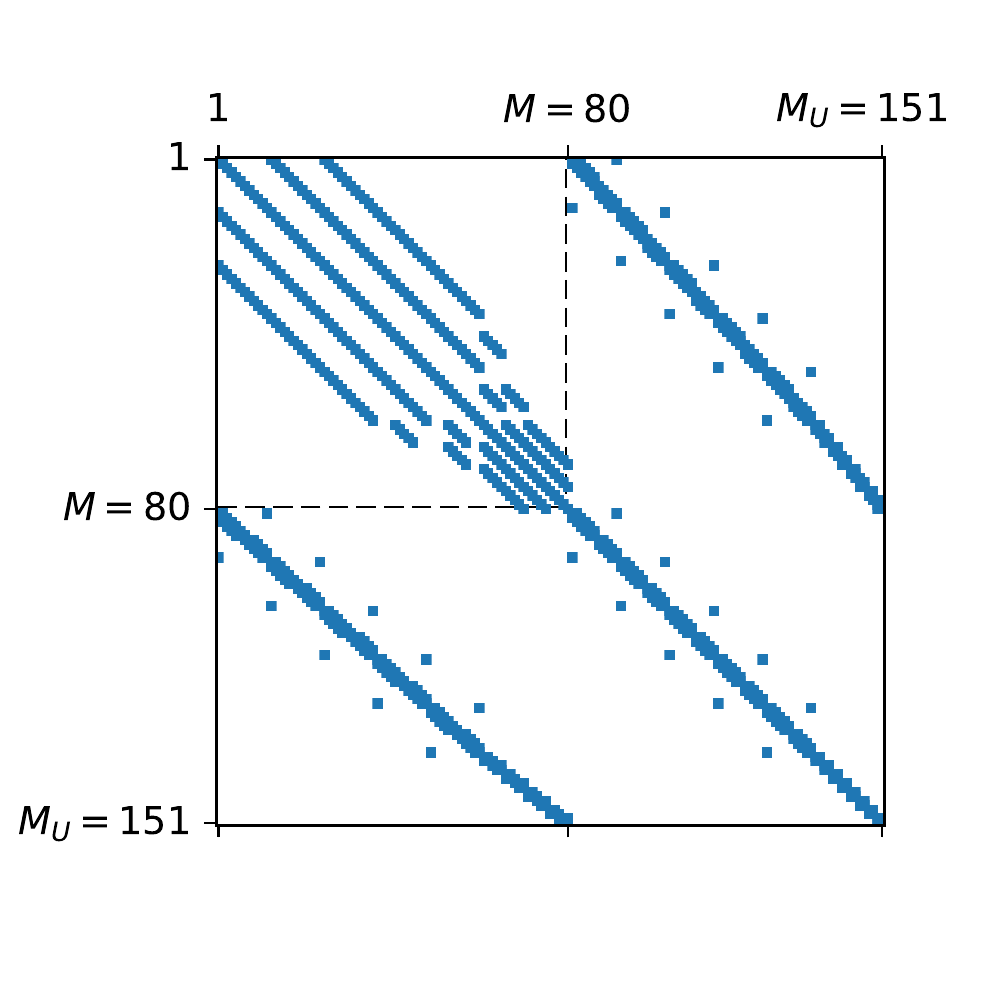}
	\caption{Patterns of the matrices associated to the algebraic linear systems emerging from the linearization of the governing equations for the {\SPeta } formulation of the problem (left) and the {\APeta } formulation of the problem (right). \label{fig:patterns}}
\end{figure}

This discretization of the differential equations, together with the shape of the computational domain, result in a sparse matrix $\mathbf{A}$ of the emerging linear system having a peculiar pattern. This pattern is represented in Fig.~\ref{fig:patterns} for both {\SPeta } and {\APeta } formulations of the problem, for a particularly poor discretization of the computational domain for the sake of the visualization ($M_z=12$, $M_r=8$, $M=80$; $M_U=80$ for the {\SPeta}, $M_U=151$ for the {\APeta } problem).

In the case of the {\SPeta } formulation, it is clearly visible that the band of the sparse matrix changes in correspondence to the change of the longitudinal size of the domain $\Omega$. This pattern is reproduced in the upper left block of the matrix pertaining to the {\APeta } formulation (in this case, the unknowns numbered from $1$ to $M$ represent the grid values of the variable $\phi$, and the unknowns ranging from $M+1$ to $M_U$ represent the grid values of the microscopic function $q$).\\

The linear system associated to this sparse matrix is solved by means of the MUMPS library \cite{MUMPS}. The error analysis, provided in Section \ref{SEC5}, is carried out by means of an estimate of the upper bound of the error affecting the solution \cite{Arioli}-- a metric which turns out to be particularly meaningful for sparse linear systems arising from a physical application -- as well as by means of the  traditional condition number \cite{Strang}. The first metric is directly provided by the MUMPS library; the second is calculated with the aid of the \texttt{linalg} module of the SciPy software package \cite{SciPy}.

\section{Simulation results}\label{SEC5}

In this section, selected numerical solutions are presented for two particular case studies. For this purpose, we have designed and implemented a code that allows to solve the model equations Eqs.~\eqref{EQev}--\eqref{BCNL} via both the {\SPeta } scheme -- leading to Eqs.~\eqref{eq:SP-discr-0}--\eqref{eq:SP-discr-1} -- and the {\APeta } scheme, leading to Eqs.~\eqref{eq:AP1-discr-0}--\eqref{eq:AP1-discr-1}.

\smallskip 

The first case, a mathematical example denoted as \textit{Case (M)}, is a set-up for which a time-independent analytical solution has been constructed. The aim of this investigation is twofold. The primary aim is to validate the developed codes by comparing the numerical solution to the exact analytical solution. 
The second main purpose is to compare the numerical solution obtained by means of the {\SPeta } formulation to the one obtained with the {\APeta } formulation, in order to highlight how the {\APeta } approach allows to overcome the difficulties encountered when using the {\SPeta } approach, as the perturbation parameter $\eta \to 0$.

\smallskip 

The second case, a physical example denoted as \textit{Case (P)}, is a set-up closer to a real physical scenario inspired by the Tokamak configuration of interest for the research group developing the TOKAM3X code. The main purpose here is to investigate if the {\APeta}-based formulation represents a viable approach to the numerical study of the problem in a practical situation.

\smallskip 

The details about the domain size, the model parameters and other settings of the numerical algorithms for the two above-mentioned cases are detailed in Section \ref{SEC51}.
In Section \ref{SEC52} and in Section \ref{SEC53}, selected numerical solutions for \textit{Case (M)} and \textit{Case (P)}, respectively, are presented.

\subsection{Presentation of the test cases}\label{SEC51}

The mathematical example, \textit{Case (M)}, regards a setting of the problem for which a time-independent exact solution was explicitly constructed. The set of model parameters ($\nu$, $\Lambda$ and $\eta$), as well as the geometric configuration (dimensions $a$, $b$, $L_z$, $l$, $L_r$; see Fig.~\ref{fig:domain}), are listed in Tab.~\ref{tab:cases}. 
\begin{table}[h]
	\begin{tabular}{ccccccccc}  
		\toprule
		 & \multicolumn{3}{c}{model parameters} & \multicolumn{5}{c}{geometric configuration} \\
		 \cmidrule(r){2-4} \cmidrule(r){5-9}
		case study  & $\eta$ & $\Gamma$ & $\nu$ & $a$ & $b$ & $L_z$ & $l$ & $L_r$ \\
		\midrule 
		\textit{Case (M)} & $\left[0,\, 1\right]$ & $1$ & $1$ & $1$ & $2$ & $3$  & $1$ & $2$ \\
		\textit{Case (P)} & $\left[0,\, 1\right]$ & $3$ & $10^{-3}$ & $1$ & $19$ & $20$  & $1$ & $2$ \\
		\bottomrule
	\end{tabular}
	\medskip 
	\caption{Model parameters and geometrical configuration of \textit{Case (M)} and \textit{Case (P)}. \label{tab:cases} }
\end{table}

The exact analytical solution $\phi$, denoted as $\phi^{(M)}_{ex}$, along with the source terms $\mathcal{S}$ and $\mathcal{F}$, denoted respectively as ${\mathcal S}^{(M)}$ and $\mathcal F^{(M)}$, are defined as follows:   
\begin{gather}
	\phi^{(M)}_{ex}(r,z) \coloneqq \sin \left(2\pi z\right) \cos \left(\pi r\right) + \Lambda + \eta \sin\left(2\pi z\right), \label{eq:caseM-exsol}\\
	{\mathcal S}^{(M)}(r,z) \coloneqq 4 \pi^2 \sin\left(2\pi z\right) + \nu \pi^4 \sin \left(2\pi z\right) \cos\left(\pi r\right), \label{eq:caseM-exsol2}\\
	{\mathcal F}^{(M)}(r,z) \coloneqq 2\pi \cos\left(2\pi z\right) \cos\left(\pi r\right) + 2\pi \eta. \label{eq:caseM-exsol3}
\end{gather}
It is worth noticing that the parameters listed in Tab.~\ref{tab:cases} as well as the source terms $\mathcal{S}^{(M)}$, $\mathcal{F}^{(M)}$ and the solution $\phi^{(M)}_{ex}$, do not have any relevant physical meaning. The only purpose of their choice is to have at our disposal a relatively simple analytical solution to Eqs.~\eqref{EQev}--\eqref{BCNL} that allows us to validate the numerical code and to investigate the good properties of the asymptotic-preserving formulation of the problem.

\medskip 

The physical example, \textit{Case (P)}, regards a more physically-oriented set-up of the problem. The model parameters and the geometric configuration -- also listed in Tab.~\ref{tab:cases} -- completely characterize the physical systems together with the source terms ${\mathcal S}$ and $\mathcal F$, denoted in this case respectively as $\mathcal{S}^{(P)}$ and $\mathcal{F}^{(P)}$, and defined as
\begin{gather}
	{\mathcal S}^{(P)}(r,z) \coloneqq 2 \cdot 10^{-3} \exp\left(-20\, L_r \left(r-\frac{3}{4} L_r\right)^2 \right), \label{eq:caseP_SS}\\ 
	{\mathcal F}^{(P)}(r,z) \coloneqq 4 \cdot 10^{-4} \cos\left(2\pi \frac{z}{L_z}\right) \exp\left( -2\, L_r \left(r-l\right)^2\right). \label{eq:caseP_SF}
\end{gather}
The analysis of the configuration \textit{Case (P)} is motivated by the purpose of applying the proposed numerical scheme to a real physical application as the study of a Tokamak plasma: in this respect, \textit{Case (P)} resembles (despite significant simplifications and idealizations) to the actual configuration and working scenario of a Tokamak plasma of interest for the studies carried out by the team developing the numerical code TOKAM3X \cite{GalassiTamain}.

\smallskip 

In both settings -- \textit{Case (M)} and \textit{Case (P)} -- the initial data $\phi_{in}\left(r,z\right) \coloneqq \phi\left(t=0,r,z\right)$ is defined as follows:
\begin{equation} \label{eq:phi0}
	\phi_{in}^{(M)}(r,z) = \phi_{in}^{(P)}(r,z) \coloneqq \Lambda.
\end{equation}

\subsection{Numerical investigations of \textit{Case (M)}}\label{SEC52}

\begin{figure}
	\includegraphics[scale=0.4]{\figdir/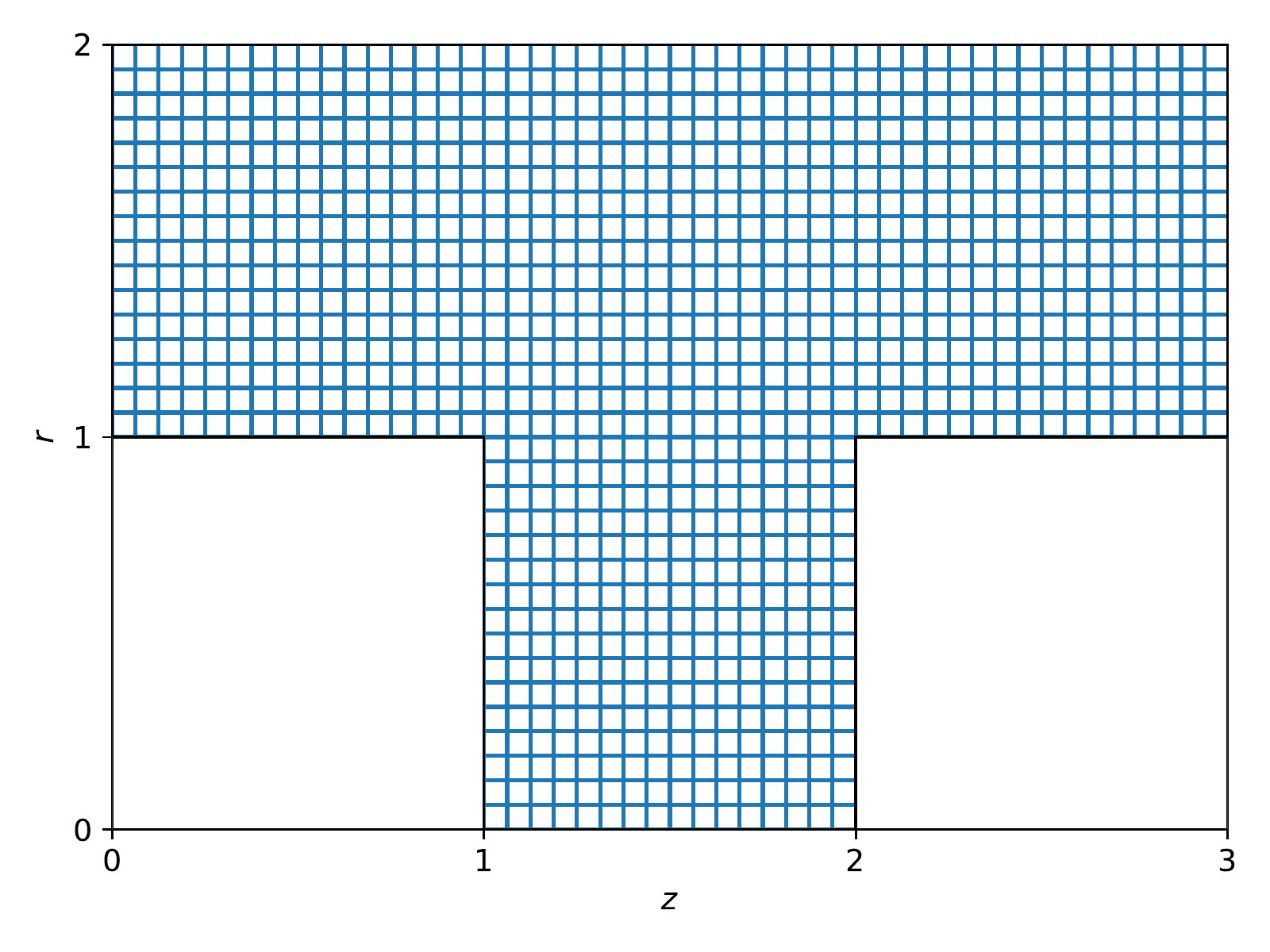}
	\caption{\textit{Case (M).} Sketch of the computational domain with an example of its discretization. For this test case, the adopted grid is uniform, with constant step size in each direction, $\Delta z = \Delta r = h$. \label{fig:caseM_grid} }
\end{figure}

Let us test now the performances of both, {\SPeta } and {\APeta } formulations, in \textit{Case (M)}. An exact steady-state solution to Eqs.~\eqref{EQev}--\eqref{BCNL} is given in \eqref{eq:caseM-exsol}. In this case, the adopted grid is uniform, with constant and equal step size in each direction, $\Delta z = \Delta r = h$; a sketch of the domain with an example of its discretization is provided in Fig.~\ref{fig:caseM_grid}.

\subsubsection{Validation of the numerical code and order of convergence}\label{SEC521}

The $L^2$-norm of the error of the numerical solution (with respect to the exact solution), as well as the order of convergence of the {\SPeta } and {\APeta } numerical schemes, are shown in Tab.~\ref{tab:caseM_L2norm}.

\begin{table}
	\begin{tabular}{lcc||cc||cc|cc}  
		\toprule
		& & & \multicolumn{2}{c||}{$\eta=0$} & \multicolumn{4}{c}{$\eta=10^{-14}$} \\
		\cmidrule(r){4-5} \cmidrule(r){6-9}
		& & & \multicolumn{2}{c||}{AP scheme} & \multicolumn{2}{c|}{SP scheme} & \multicolumn{2}{c}{AP scheme} \\
		\cmidrule(r){4-5} \cmidrule(r){6-7} \cmidrule(r){8-9}
		$h$    & $M_z \times M_r$ & $M$ & $L^2$-norm error& $p$ & $L^2$-norm error& $p$ & $L^2$-norm error& $p$ \\
		\midrule
		$1/4$   & $ 12 \times   9$ & $      80$ & $8.0960 \times 10^{-1}$ & $2.258$ & $1.3123 \times 10^{-1}$  & $2.119$ & $8.0960 \times 10^{-1}$ & $2.258$ \\
		$1/8$   & $ 24 \times  17$ & $     288$ & $1.6925 \times 10^{-1}$ & $2.051$ & $3.0213 \times 10^{-1}$  &     $-$ & $1.6925 \times 10^{-1}$ & $2.051$ \\
		$1/16$  & $ 48 \times  33$ & $  1\,088$ & $4.0852 \times 10^{-2}$ & $2.013$ &                \checkno  &     $-$ & $4.0852 \times 10^{-2}$ & $2.013$ \\
		$1/32$  & $ 96 \times  65$ & $  4\,224$ & $1.0123 \times 10^{-2}$ & $2.003$ &                \checkno  &     $-$ & $1.0123 \times 10^{-2}$ & $2.003$ \\
		$1/64$  & $192 \times 129$ & $ 16\,640$ & $2.5251 \times 10^{-3}$ & $1.995$ &                \checkno  &     $-$ & $2.5252 \times 10^{-3}$ & $1.989$ \\
		$1/128$ & $384 \times 257$ & $ 66\,048$ & $6.3342 \times 10^{-4}$ & $1.999$ &                \checkno  &     $-$ & $6.4528 \times 10^{-4}$ & $1.976$ \\
		$1/256$ & $768 \times 513$ & $263\,168$ & $1.5852 \times 10^{-4}$ &     $-$ &                \checkno  &     $-$ & $1.6403 \times 10^{-4}$ &     $-$ \\
		\bottomrule
	\end{tabular}
	\medskip 
	\begin{tabular}{l||cc|cc||cc|cc}  
		\toprule
		& \multicolumn{4}{c||}{$\eta=10^{-8}$} & \multicolumn{4}{c}{$\eta=1$} \\
		\cmidrule(r){2-5} \cmidrule(r){6-9}
		& \multicolumn{2}{c|}{SP scheme} & \multicolumn{2}{c||}{AP scheme} & \multicolumn{2}{c|}{SP scheme} & \multicolumn{2}{c}{AP scheme}\\
		\cmidrule(r){2-3} \cmidrule(r){4-5} \cmidrule(r){6-7} \cmidrule(r){8-9}
		$h$    & $L^2$-norm error& $p$ & $L^2$-norm error& $p$ & $L^2$-norm error& $p$ & $L^2$-norm error& $p$ \\
		\midrule
		$1/4$   & $8.0960 \times 10^{-1}$  & $2.258$ & $8.0960 \times 10^{-1}$ & $2.258$ & $7.1804 \times 10^{-1}$  & $2.200$ & $7.1804 \times 10^{-1}$ & $2.200$\\
		$1/8$   & $1.6925 \times 10^{-1}$  & $2.051$ & $1.6925 \times 10^{-1}$ & $2.051$ & $1.5626 \times 10^{-1}$  & $2.033$ & $1.5626 \times 10^{-1}$ & $2.033$\\
		$1/16$  & $4.0852 \times 10^{-2}$  & $2.013$ & $4.0852 \times 10^{-2}$ & $2.013$ & $3.8172 \times 10^{-2}$  & $2.007$ & $3.8172 \times 10^{-2}$ & $2.007$\\
		$1/32$  & $1.0124 \times 10^{-2}$  & $1.984$ & $1.0123 \times 10^{-2}$ & $1.995$ & $9.4944 \times 10^{-3}$  & $2.003$ & $9.4944 \times 10^{-3}$ & $2.003$\\
		$1/64$  & $2.5586 \times 10^{-3}$  & $1.821$ & $2.5387 \times 10^{-3}$ & $1.968$ & $2.3689 \times 10^{-3}$  & $2.001$ & $2.3689 \times 10^{-3}$ & $2.001$\\
		$1/128$ & $7.2401 \times 10^{-4}$  &     $-$ & $6.4899 \times 10^{-4}$ & $2.005$ & $5.9181 \times 10^{-4}$  & $1.980$ & $5.9161 \times 10^{-4}$ & $1.953$\\
		$1/256$ &                 \checkno &     $-$ & $1.6172 \times 10^{-4}$ &     $-$ & $1.4997 \times 10^{-4}$  &     $-$ & $1.5282 \times 10^{-4}$ &     $-$\\
		\bottomrule
	\end{tabular}
	\caption{\textit{Case (M).} $L^2$-norm of the error of the numerical solution with respect to the exact solution, $\norm{\phi^{\eta} - \phi^{(M)}_{ex}}_{2}$, and estimated order of convergence, $p$, of the {\APeta } and {\SPeta } schemes, for several grid sizes ($\Delta z = \Delta r = h$) and $\eta=0,\, 10^{-14},\, 10^{-8},\, 1$. The symbol {\checkno } denotes the lack of a numerical solution due to blow-up of the numerical scheme. \label{tab:caseM_L2norm}}
\end{table}

It is evident that the {\APeta } numerical scheme guarantees good performances for any value of the parameter $\eta$ in the interval $[0,\,1]$; even values of $\eta$ as small as $10^{-14}$ or the limit case $\eta\equiv 0$ are not critical. Contrary to this, the {\SPeta } scheme can be used only for larger values of $\eta$. As Tab.~\ref{tab:caseM_L2norm} shows, the case $\eta = 10^{-14}$ is only treatable with the {\APeta } scheme, and even in the case $\eta = 10^{-8}$, the {\SPeta } scheme does not allow to obtain reasonable numerical results (or any result at all) when the grid is fine. 

\smallskip

\begin{figure}
	\includegraphics[scale=0.6]{\figdir/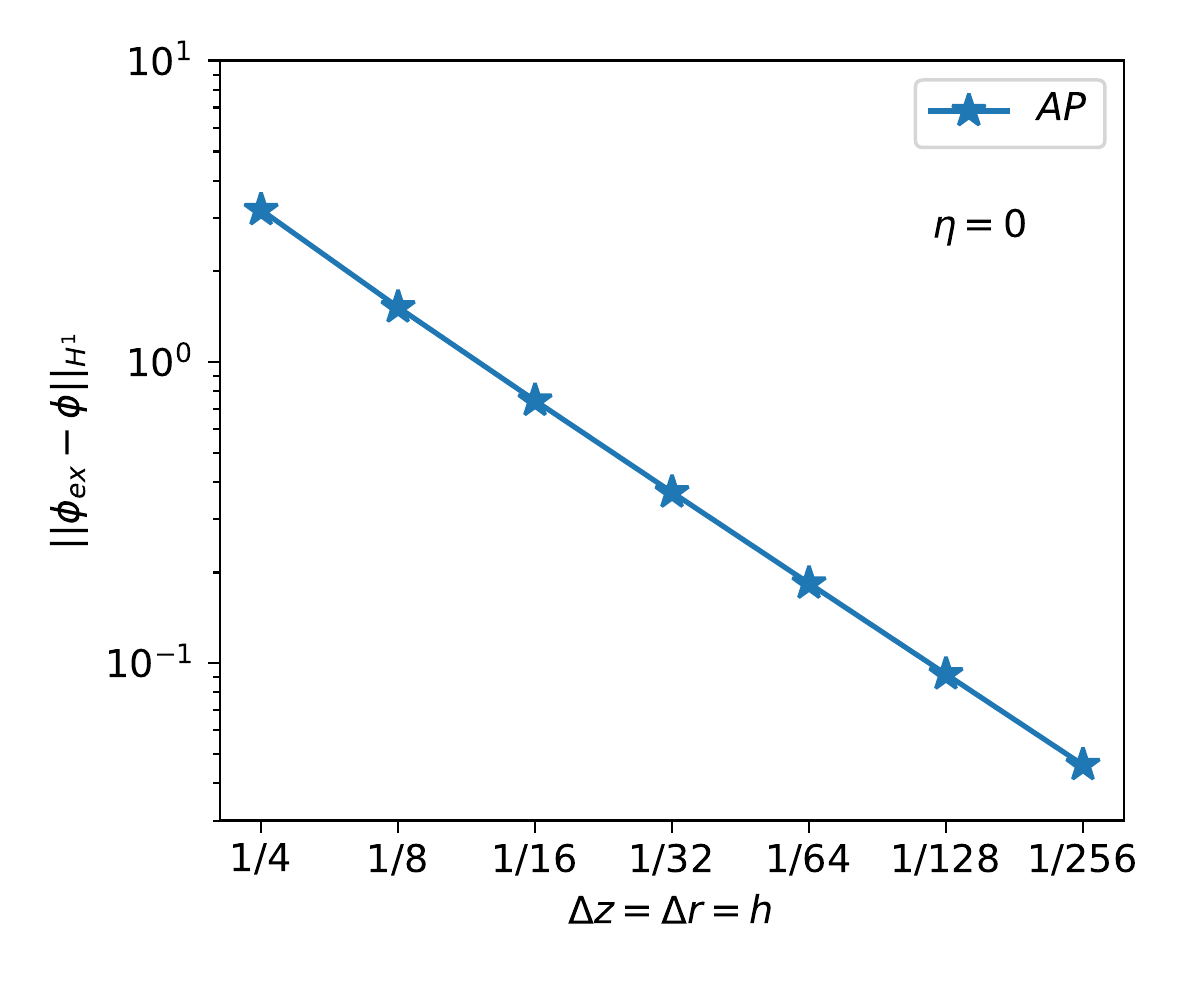}
	\includegraphics[scale=0.6]{\figdir/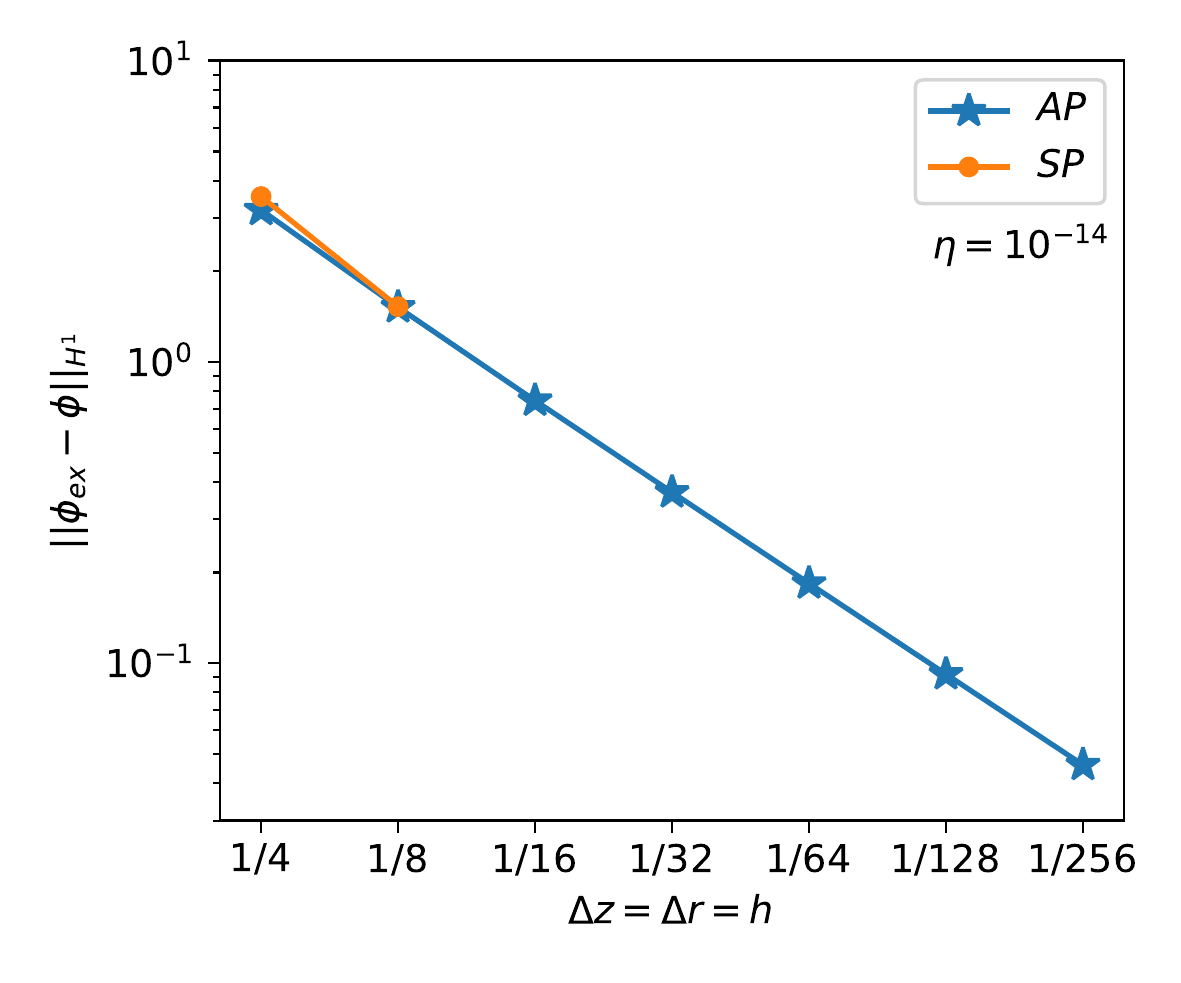} \\
	\includegraphics[scale=0.6]{\figdir/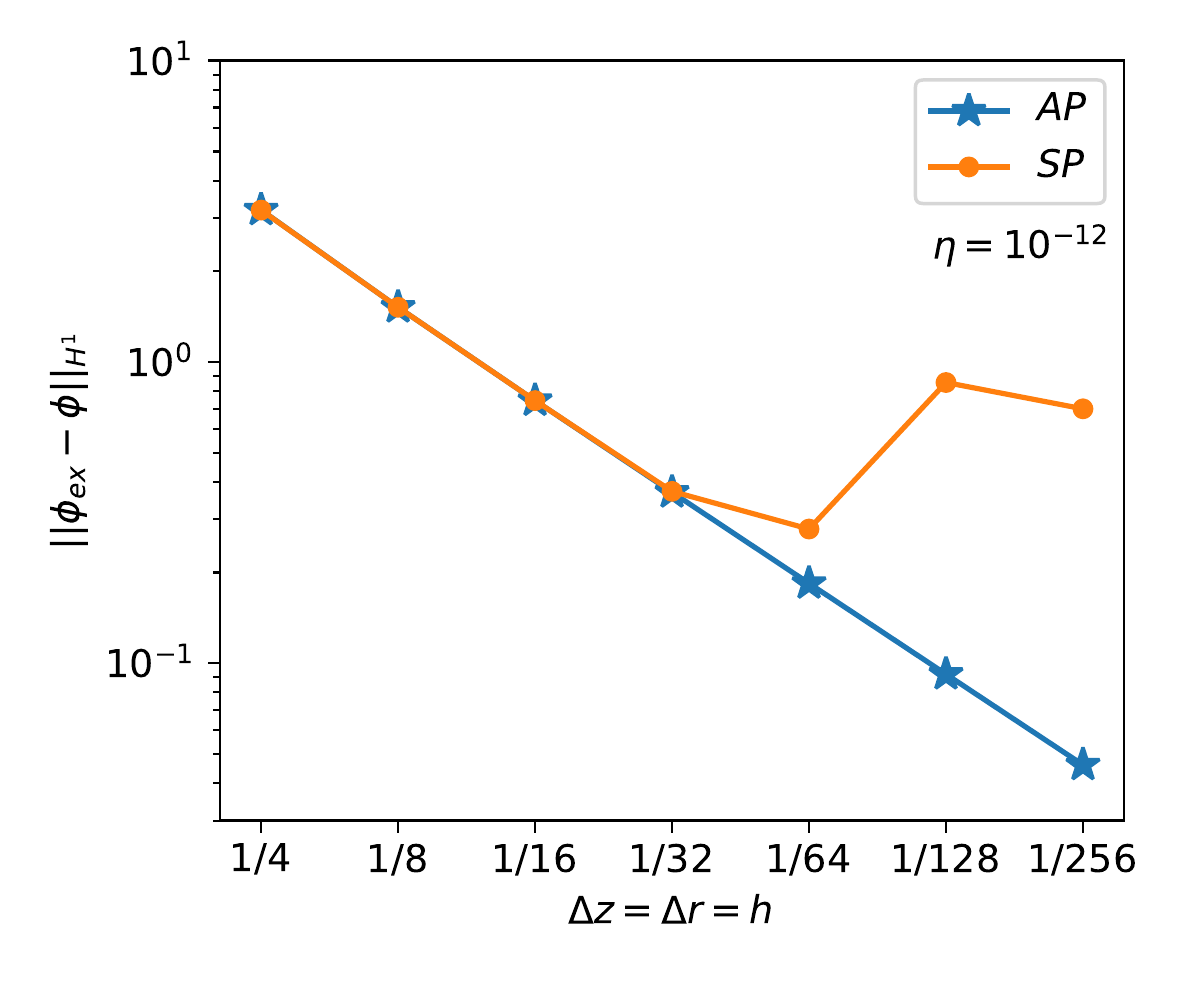}
	\includegraphics[scale=0.6]{\figdir/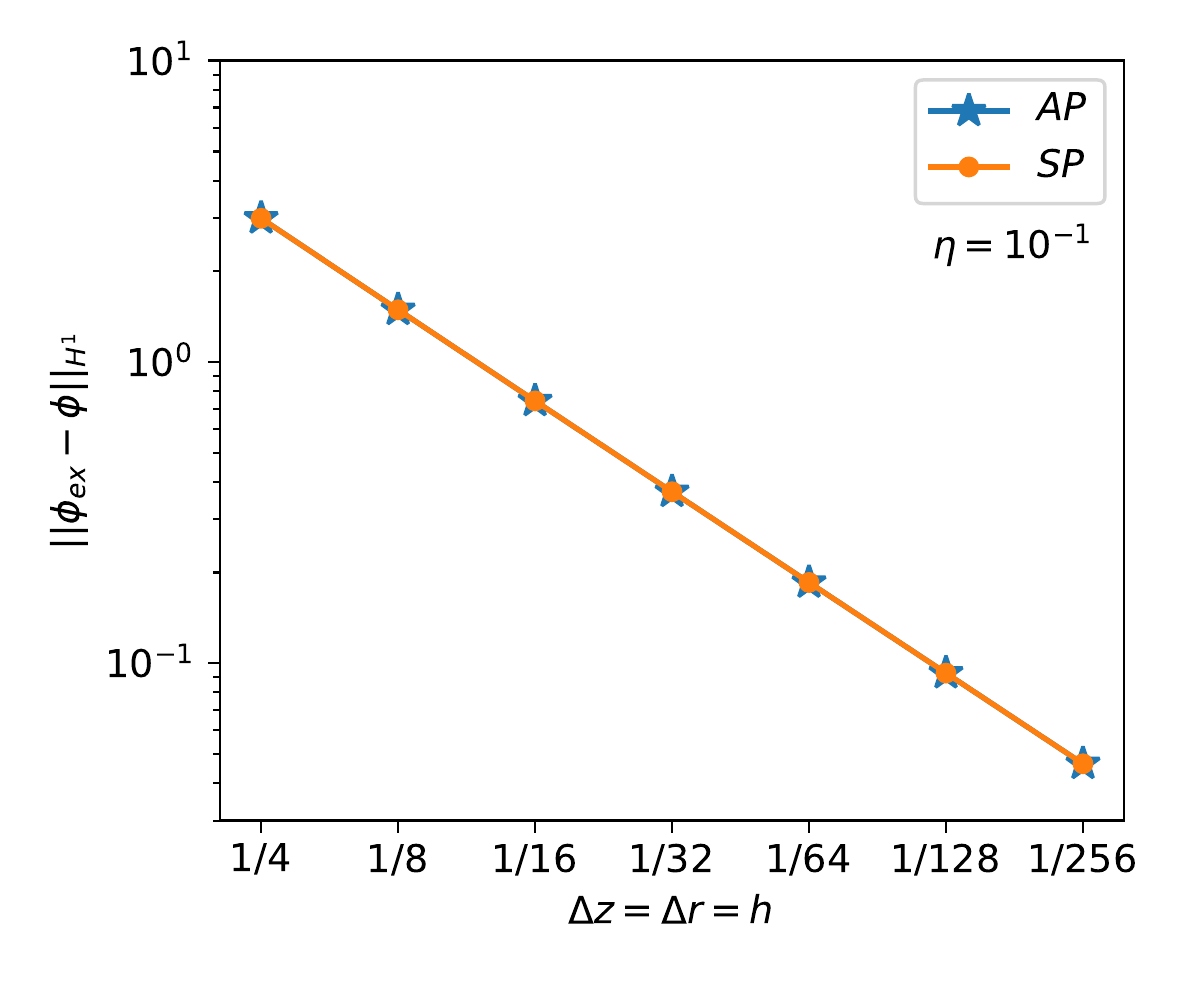}
	\caption{\textit{Case (M).} $H^1$-norm of the error of the numerical solution with respect to the exact solution, $\norm{\phi^{\eta} - \phi^{(M)}_{ex}}_{H^1}$, as a function of the grid size and for several $\eta$-values. The slopes are approx. $1$. Note that for $\eta=0$ and $10^{-14}$ the {\SPeta } scheme does not furnish reasonable results.
	\label{fig:caseM_H1norm} }
\end{figure}

The behavior of the $H^1$-norm of the error of the numerical solution (with respect to the exact solution) is shown in Fig.~\ref{fig:caseM_H1norm}. Comparing these data to those shown in Tab.~\ref{tab:caseM_L2norm}, it is easily seen that when considering the $L^2$-norm, the convergence is of second order in space, while it is of first order in the $H^1$-norm, which is completely standard.

\subsubsection{Comparison of the {\SPeta } and {\APeta } formulations} \label{SEC522}

In order to compare the behavior of the {\SPeta } and {\APeta } schemes for small values of $\eta$, it is useful to take a closer look at the linear system of algebraic equations emerging from the discretization of the model equations.

\smallskip 

A classical metric which provides insight in the reliability and accuracy of the solution of a linear system $\mathbf{A} \mathbf{x} = \mathbf{b}$, is the condition number of the matrix $\mathbf{A}$, defined as $\kappa\left( \mathbf{A} \right) \coloneqq \norm{ \mathbf{A}^{-1} } \norm{\mathbf{A}}$.
The larger the condition number, the closer to singular is the matrix. In this latter case, obtaining the solution to the linear system could not be feasible or, when feasible, the solution is typically affected by a large error and therefore should not be trusted as a reliable solution.\\ 

In more details, the relative effect of round-off errors on the solution of a linear system is bounded by the condition number of the matrix $\kappa_2\left( \mathbf{A} \right) = \norm{ \mathbf{A}^{-1} }_{2} \norm{\mathbf{A}}_{2}$  in the following manner
$$
{||\delta \mathbf{x}||_2 \over || \mathbf{x}||_2} \le \kappa_2( \mathbf{A}) \left({||\delta \mathbf{A}||_2 \over || \mathbf{A}||_2}+{||\delta \mathbf{b}||_2 \over || \mathbf{b}||_2} \right)\,,
$$
where $\mathbf{\tilde{x}}=\mathbf{x}+\delta \mathbf{x}$ is the computed numerical solution of $\mathbf{A} \mathbf{x} = \mathbf{b}$ and can be seen as the exact solution of the slightly perturbed linear system $( \mathbf{A}+\delta \mathbf{A})\,\mathbf{\tilde{x}} =\mathbf{b}+\delta \mathbf{b}$, perturbations being due to round-off errors.\\
This traditional relative error-estimate is very pessimistic, for example it does not take into account for the particular form of the matrix $\mathbf{A}$, or for the special form of the right-hand side $\mathbf{b}$. A more satisfying approach has been proposed by Arioli et al. in \cite{Arioli}, where the sparsity of the matrix is kept in mind. The relative error of the solution to the linear system is estimated in that work as follows
\begin{equation*}
	\frac{ \norm{ \delta \mathbf{x} }_{\infty} }{ \norm{\mathbf{x}}_{\infty}  } \leq \bar\eps, \qquad 
	\bar\eps \coloneqq \omega_1 \kappa_{\omega_1} + \omega_2 \kappa_{\omega_2}\,,
\end{equation*}
where the quantities $\omega_1$ and $\omega_2$ (backward errors) are defined as follows
\begin{equation}
\omega_1 = \max_{i \in I_1} \left( \frac{ \abs{\mathbf{A} \mathbf{\tilde{x}} - \mathbf{b}}_i }{ \left( \abs{\mathbf{A}} \abs{\mathbf{\tilde{x}}} + \abs{\mathbf{b}} \right)_i } \right), \qquad
\omega_2 = \max_{i \in I_2} \left( \frac{ \abs{\mathbf{A} \mathbf{\tilde{x}} - \mathbf{b} }_i }{ \left( \abs{\mathbf{A}} \abs{\mathbf{\tilde{x}}} \right)_i + \norm{ \mathbf{A}_i }_{\infty} \norm{\mathbf{\tilde{x}}}_{\infty} } \right),
\end{equation}
with $\mathbf{A}_i$ the $i$-th row of $\mathbf{A}$, $\abs{\mathbf{A}}$ being the matrix with elements $\abs{A_{ij}}$, $\abs{\mathbf{b}}$ the vector with elements $\abs{b_i}$, and $I_2$ representing the set of indices of the equations of the linear system such that $\abs{\mathbf{A} \mathbf{\tilde{x}} - \mathbf{b}}_i$ is nonzero and $\left( \abs{\mathbf{A}} \abs{\mathbf{\tilde{x}}} + \abs{\mathbf{b}} \right)_i$ is small, whereas $I_1$ representing the set of indices of the remaining equations (if $I_2$ is empty, then $\omega_2=0$).

Moreover, two condition numbers $\kappa_{\omega_1}$ resp. $\kappa_{\omega_2}$  of the system (not just of the matrix) are defined as follows:
\begin{equation*}
	\kappa_{\omega_1} = \frac{ \norm{ \abs{\mathbf{A}^{-1}} \left( \abs{\mathbf{A}} \abs{\mathbf{\tilde{x}}} + \abs{\mathbf{b}} \right) }_{\infty} }{ \norm{\mathbf{\tilde{x}}}_{\infty} }, \qquad 
	\kappa_{\omega_2} = \frac{ \norm{ \abs{\mathbf{A}^{-1}} \left( \abs{\mathbf{A}} \abs{\mathbf{\tilde{x}}} + \abs{\mathbf{A}} \mathbf{1} \norm{\mathbf{\tilde{x}}}_{\infty} \right) }_{\infty} }{ \norm{\mathbf{\tilde{x}}}_{\infty} },
\end{equation*}
where $\mathbf{1}$ is the column vector with all unitary elements. In the evaluation of $\kappa_{\omega_\iota}$ only those equations with index $i\in I_\iota$ are considered ($\iota=1,2$).\\

\begin{figure}
	\includegraphics[scale=0.6]{\figdir/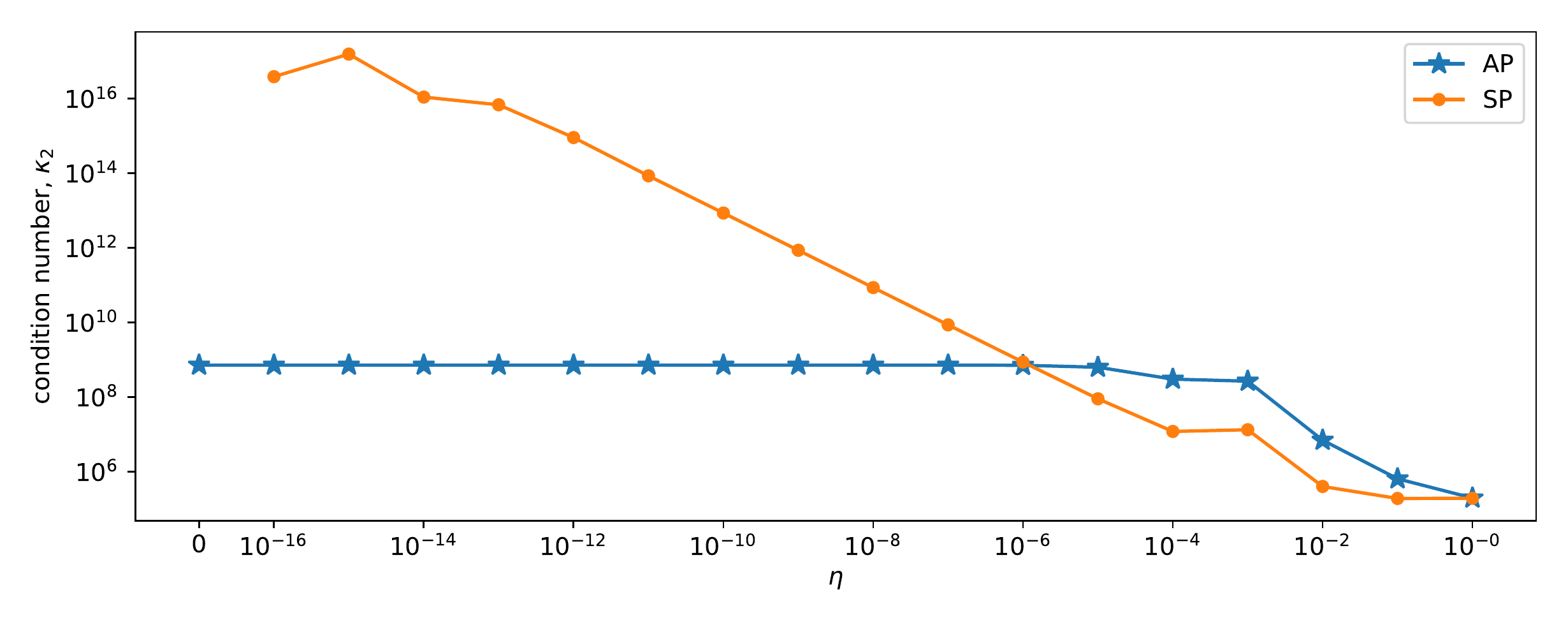}
	\includegraphics[scale=0.6]{\figdir/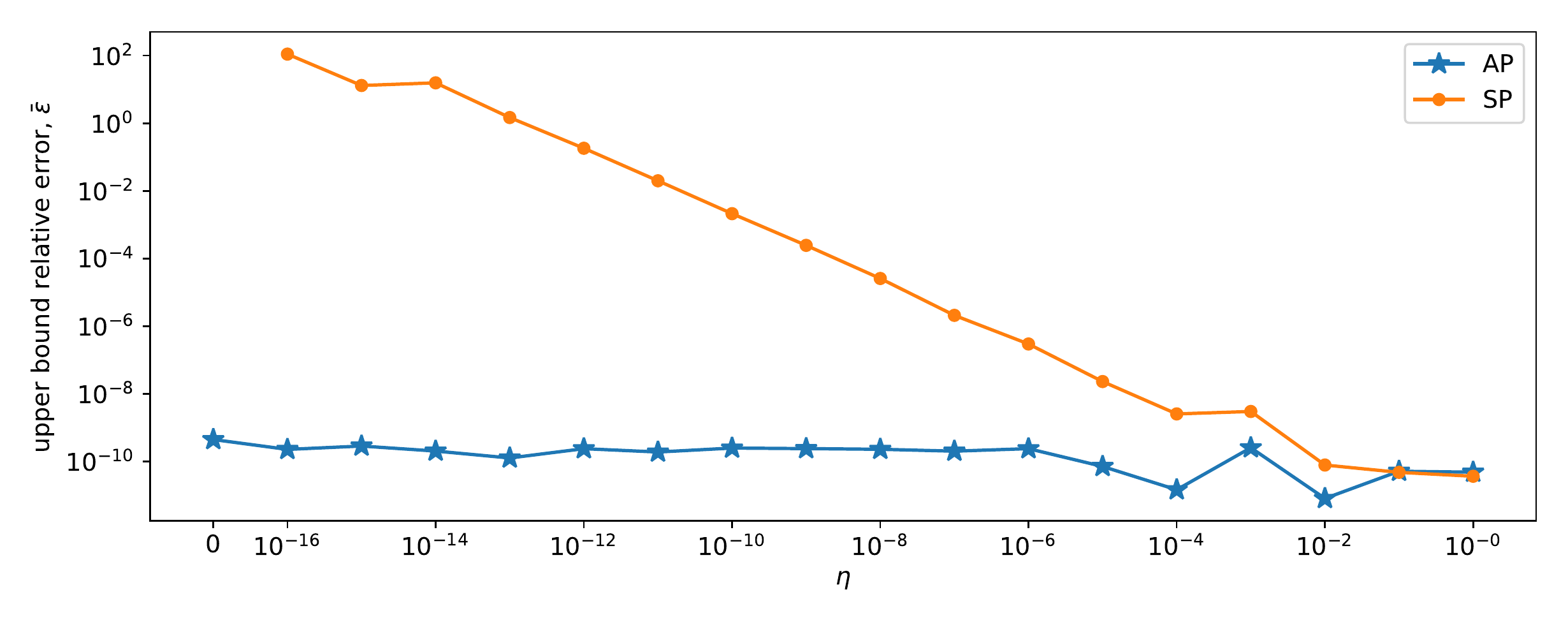}
	\caption{\textit{Case (M).} Condition number $\kappa_2$ (top) and estimate of the upper bound of the relative error $\bar\eps$ (bottom) of the lin. syst. sol., as a function of the parameter $\eta$, for both {\SPeta } and {\APeta } formulations ($M_z=24$, $M_r=17$, $M=288$; $M_U=288$ for {\SPeta }, $M_U=559$ for {\APeta }; $\Delta z = \Delta r = 1/8$). \label{fig:caseM_cond} }
\end{figure}

In order to evaluate numerically the accuracy of the resolution of the linear system corresponding to our two problems {\SPeta } and {\APeta }, in particular to examine the influence of the parameter $\eta$ on the obtained results, we plotted in Fig.~\ref{fig:caseM_cond} two graphs. Firstly, the standard condition number $\kappa_2\left( \mathbf{A} \right) = \norm{ \mathbf{A}^{-1} }_{2} \norm{\mathbf{A}}_{2}$ of the matrix $\mathbf{A}$ associated to the {\SPeta } and {\APeta } schemes, is plotted as a function of the perturbation parameter $\eta$. 
As expected, the results show that as the perturbation parameter $\eta \to 0$, the condition number $\kappa_2$ associated to the {\SPeta } increases as $\eta^{-1}$, signifying that the matrix $\mathbf{A}$ gets closer and closer to a singular-matrix. In contrast, the condition number $\kappa_2$ associated to the {\APeta } scheme never reaches critical values and remains independent of the parameter $\eta$, clearly showing that the singularity of $\eta\to 0$ of the singularly perturbed problem \eqref{EQev} has been removed in the {\APeta } formulation we propose.

\medskip 

Secondly, to be sure that the traditional condition number is not too pessimistic in the here treated case, we decided to consider also the new metric proposed in \cite{Arioli}, allowing to compute a good estimate of the upper bound of the relative error of the computed solution $\mathbf{\tilde{x}}$ with respect to the exact solution $\mathbf{x}$ of $\mathbf{A} \mathbf{x} = \mathbf{b}$.
The analysis of the upper bound $\bar\eps$ of the relative error of the computed solution, shown in Fig.~\ref{fig:caseM_cond}, once more clearly points out how the solution corresponding to the {\SPeta}-scheme rapidly becomes unreliable as $\eta \to 0$. On the other hand, the relative error affecting the solution corresponding to the {\APeta}-scheme is roughly constant and of the order $10^{-10}$--$10^{-8}$ also for $\eta \ll 10^{-6}$.
\subsection{Numerical investigations of \textit{Case (P)} } \label{SEC53}

\begin{figure}
	\includegraphics[scale=0.6]{\figdir/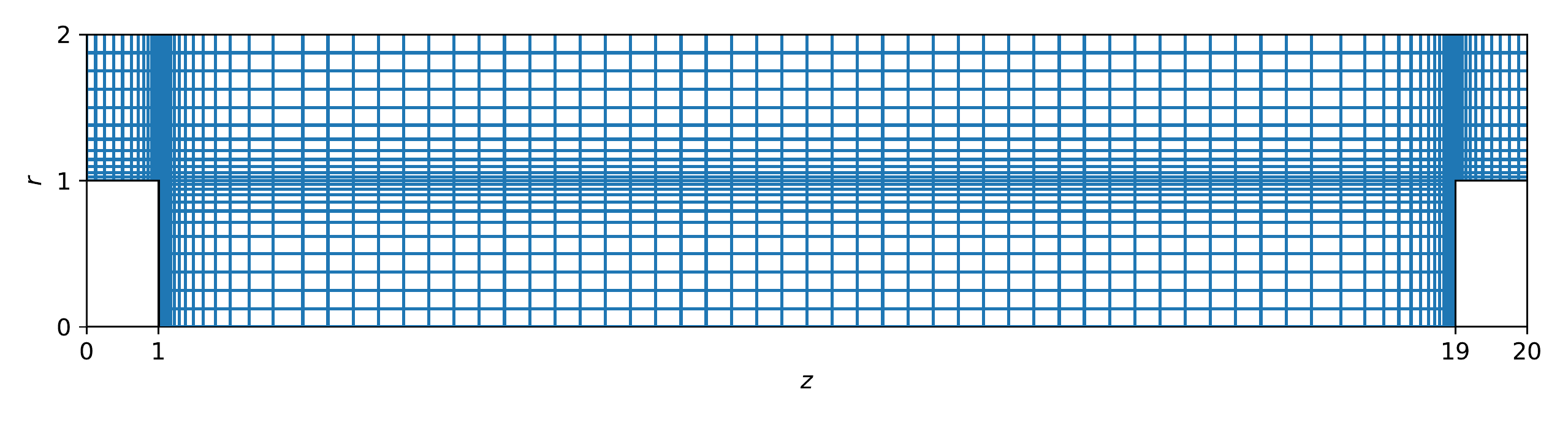}
	\caption{\textit{Case (P).} Sketch of the computational domain with an example of its discretization. For this test case, the adopted grid has variable steps $\Delta z$, $\Delta r$ in both longitudinal and radial direction, being more refined in the proximity of the domain borders $\Sigma_l$, $\Gamma_a$, $\Gamma_b$. \label{fig:caseP_grid} }
\end{figure}

\begin{figure}
	\includegraphics[scale=0.6]{\figdir/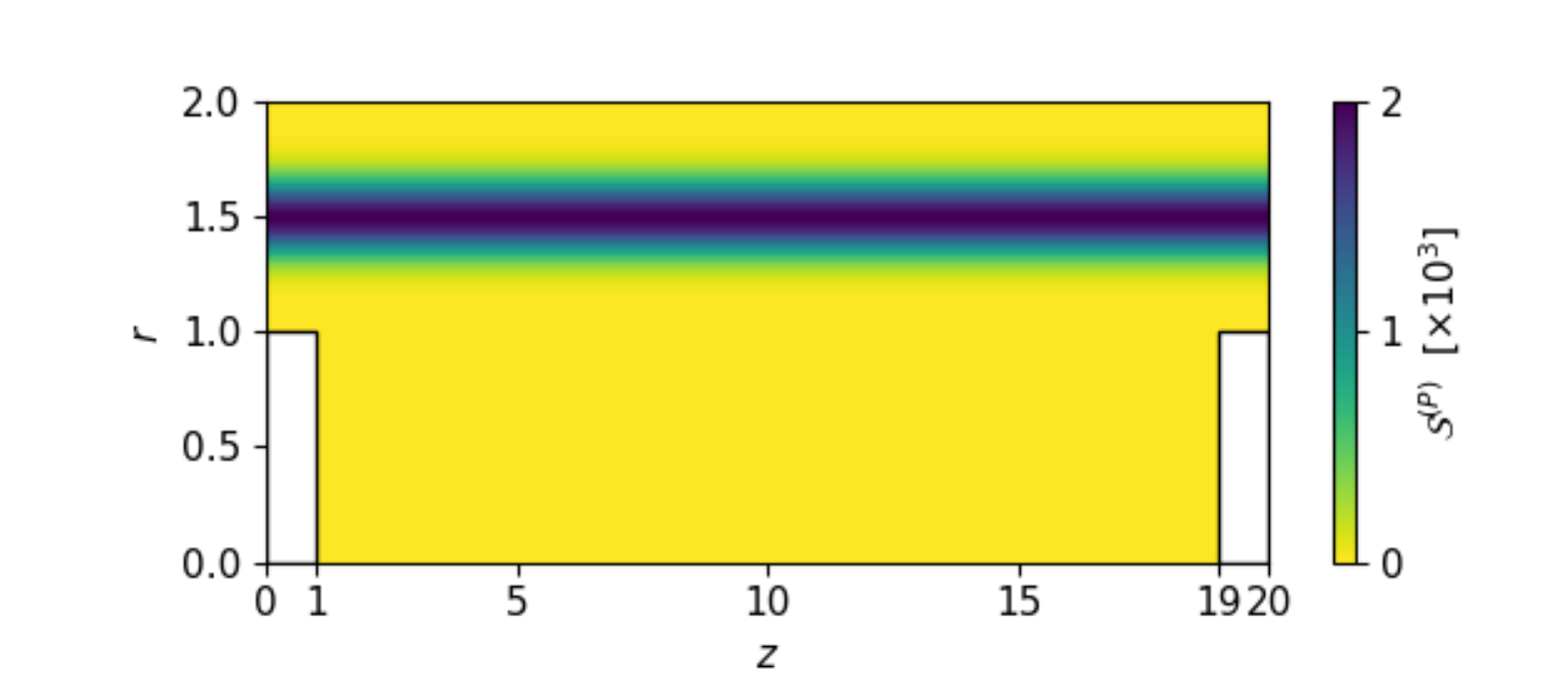} %
	\includegraphics[scale=0.6]{\figdir/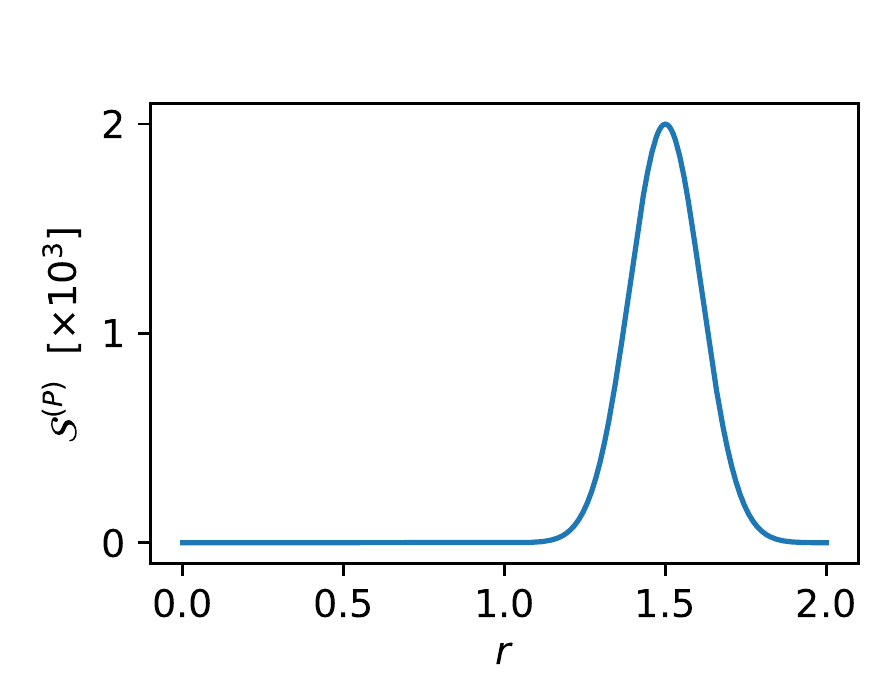}
	\includegraphics[scale=0.6]{\figdir/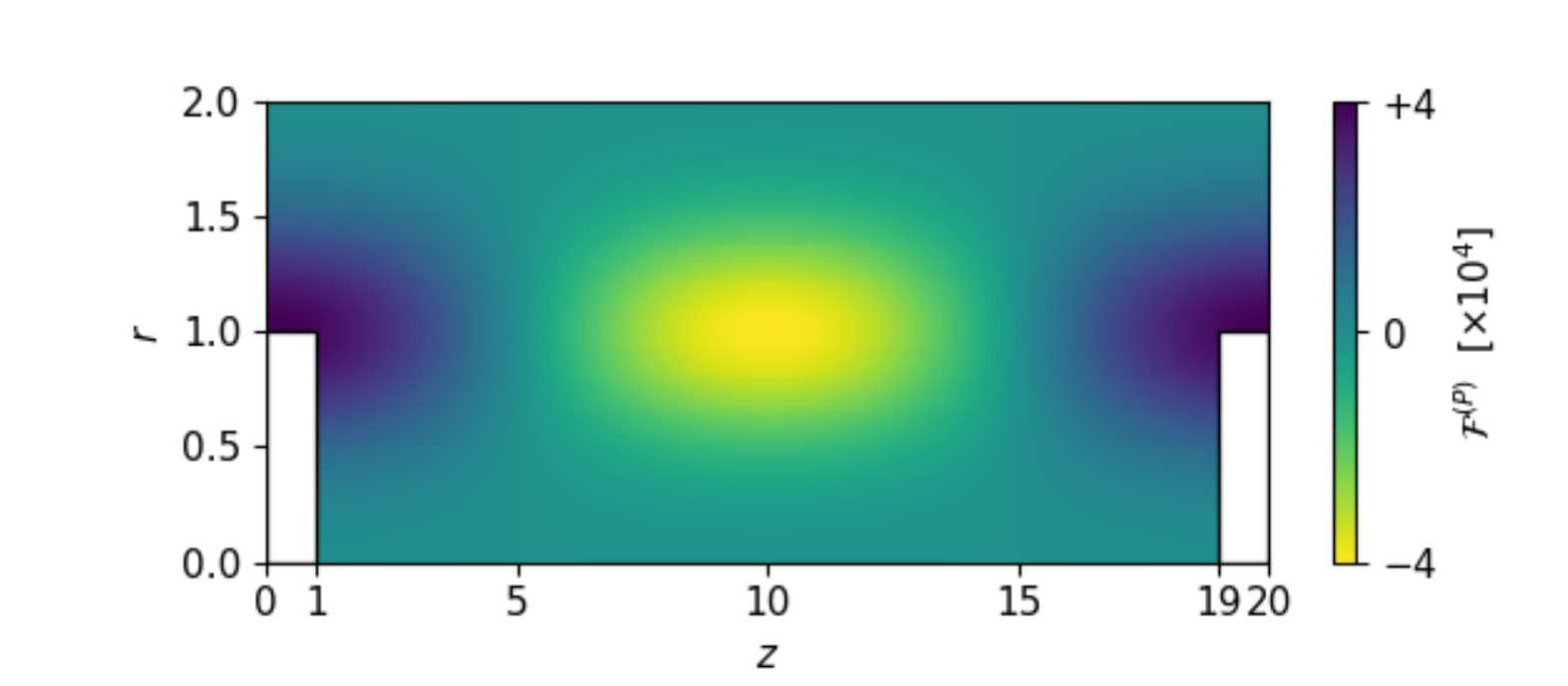} %
	\includegraphics[scale=0.6]{\figdir/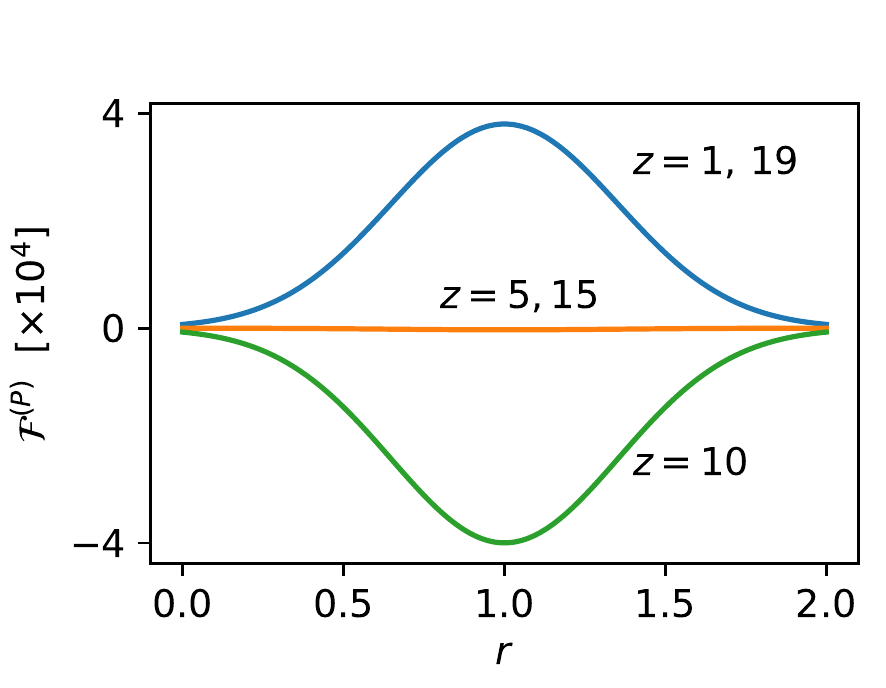}
	\caption{\textit{Case (P).} Graphical representation of the source terms $\mathcal{S}^{(P)}$ (top) and $\mathcal{F}^{(P)}$ (bottom) given in Eq.~\eqref{eq:caseP_SS}--\eqref{eq:caseP_SF}. The dependence on $r$ of $\mathcal{S}^{(P)}$ (which is constant in $z$ direction), and the dependence on $r$ of $\mathcal{F}^{(P)}$ at three different values of $z$ ($z=1,\, 5,\, 10$) are plotted as well. \label{fig:caseP_sourcesSF} }
\end{figure}

In this section, we discuss the numerical results obtained for a case inspired by a real physical application, namely the evolution of the electric potential in the peripheral plasma region of a tokamak reactor.
The list of model parameters and the geometric configuration of this case are given in Tab.~\ref{tab:cases}, while a sketch of the discretized computational domain is provided in Fig.~\ref{fig:caseP_grid}. In contrast to the mathematical case previously discussed in Section \ref{SEC52}, the computational grid is not uniform in this case, being more refined (i.e. with smaller step sizes $\Delta z$ and $\Delta r$) in the proximity of the borders $\Gamma_l$, $\Gamma_a$, $\Gamma_b$. 
A graphical representation of the source terms $\mathcal{S}^{(P)}$ and $\mathcal{F}^{(P)}$, given in Eqs.~\eqref{eq:caseP_SS}--\eqref{eq:caseP_SF}, is provided in Fig.~\ref{fig:caseP_sourcesSF}.

\subsubsection{Error estimates of the numerical solution}\label{SEC531}
To start, we computed also in this physical case the condition number $\kappa_2$ and the upper bound of the relative error $\bar\eps$, introduced in Section \ref{SEC522}, for the {\APeta } scheme and compared it to those obtained for the {\SPeta } scheme.

\begin{figure}
	\includegraphics[scale=0.6]{\figdir/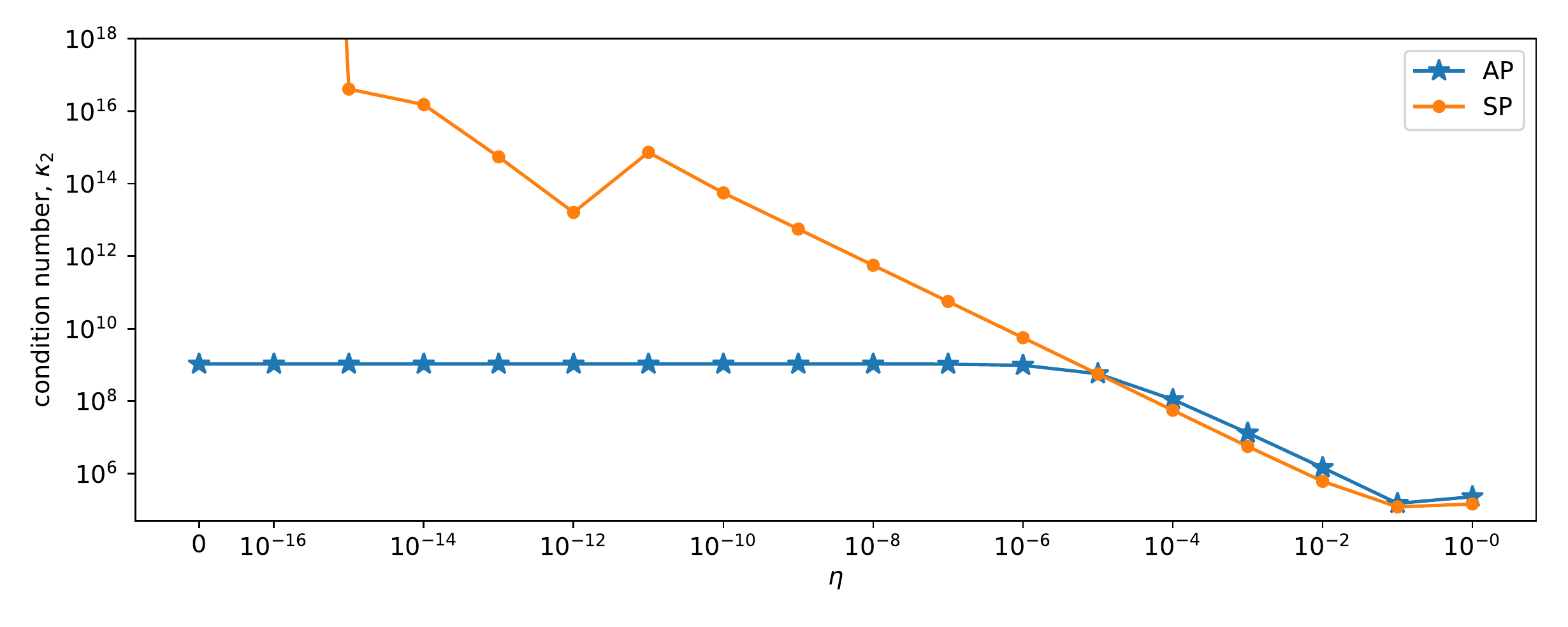}
	\includegraphics[scale=0.6]{\figdir/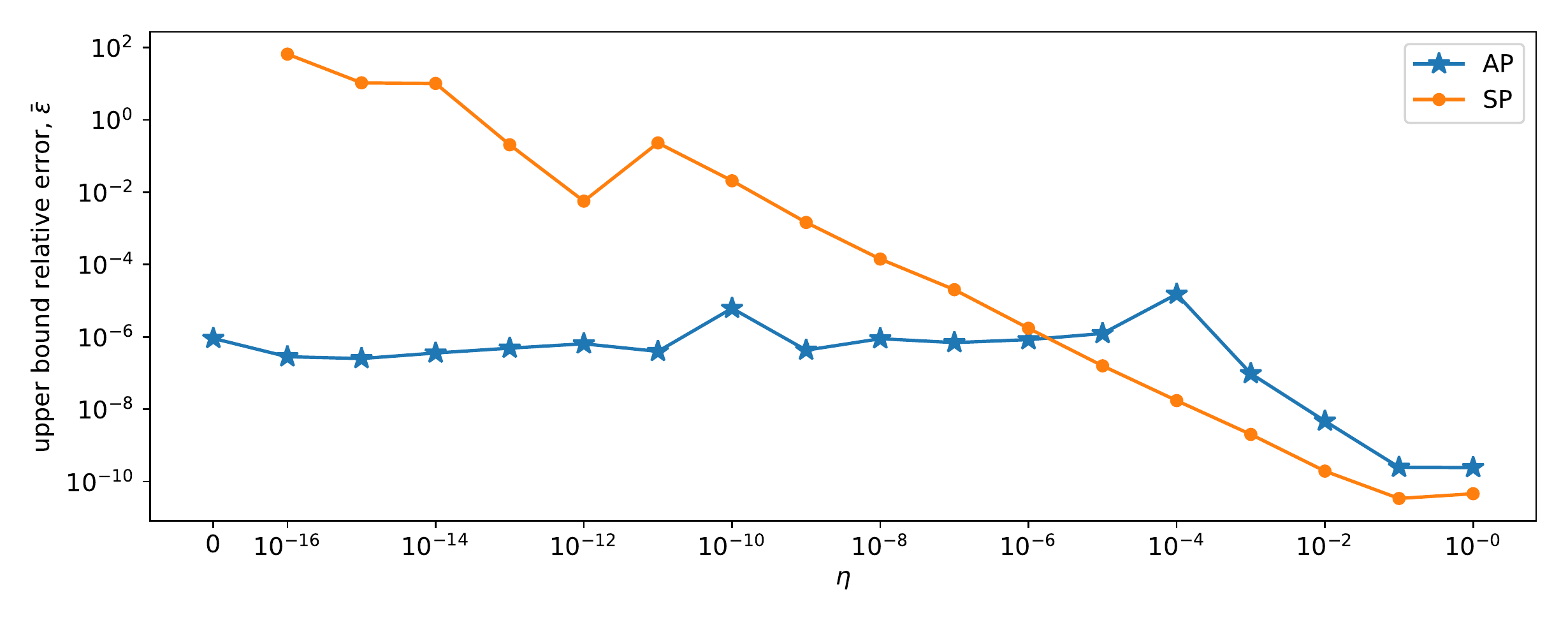}
	\caption{\textit{Case (P).} Condition number $\kappa_2$ (top) and estimate of the upper bound of the relative error $\bar\eps$ (bottom) of the lin. syst. sol. as a function of the parameter $\eta$, for both {\SPeta } and {\APeta } formulations ($M_z=160$, $M_r=17$, $M=2\,600$; $M_U=2\,600$ for {\SPeta}, $M_U=5\,1853$ for {\APeta}). \label{fig:caseP_cond} }
\end{figure}

These metrics, plotted in Fig.~\ref{fig:caseP_cond}, clearly show that -- as expected -- also in this physical case the {\SPeta } scheme does not provide reliable numerical results for values of the parameter $\eta$ below $10^{-8}$, while the {\APeta }-scheme always gives accurate numerical results independently on the parameter $\eta$. The results are naturally obtained with a fixed grid and $\eta$ varying in $[0,1]$, clearly underlying the Asymptotic-Preserving property of the numerical scheme we introduced in this work.

\subsubsection{Selection of numerical results}\label{SEC532}
Let us now study in more details the numerical results obtained for the physical \textit{Case (P)}.

\smallskip

\begin{figure}
	\includegraphics[scale=0.6]{\figdir/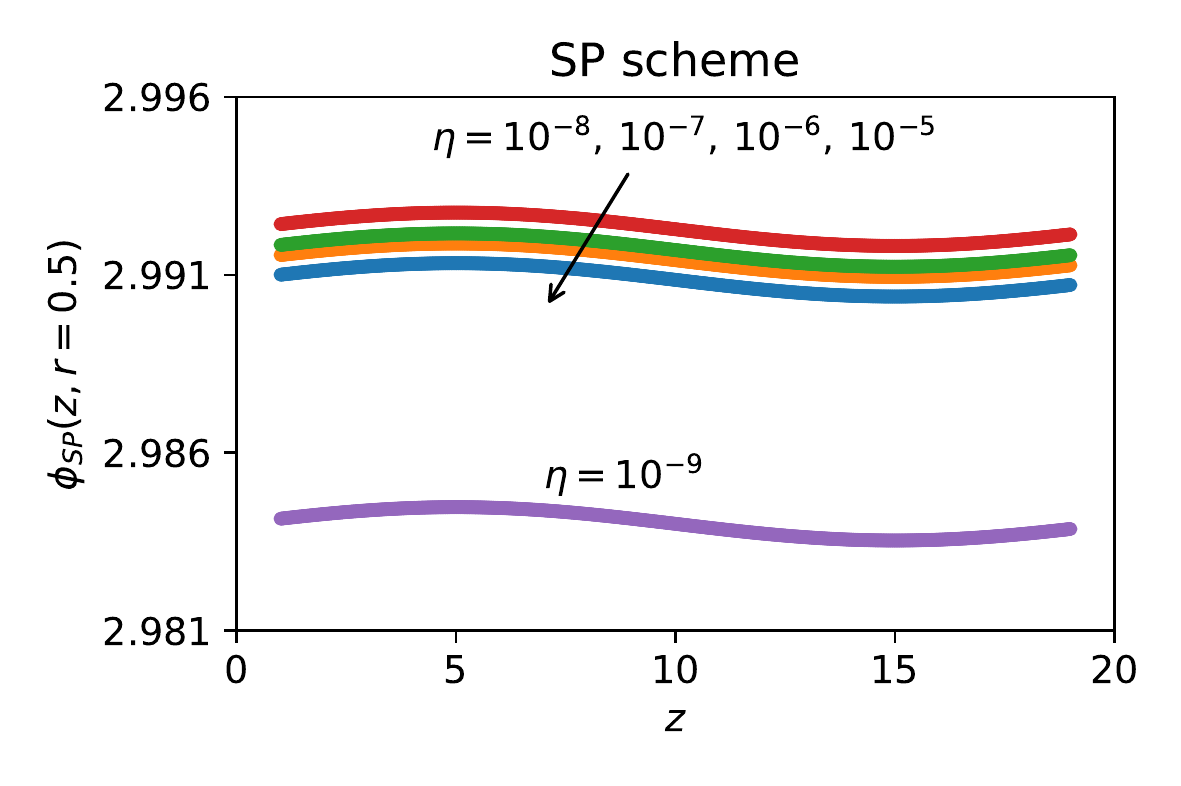}
	\includegraphics[scale=0.6]{\figdir/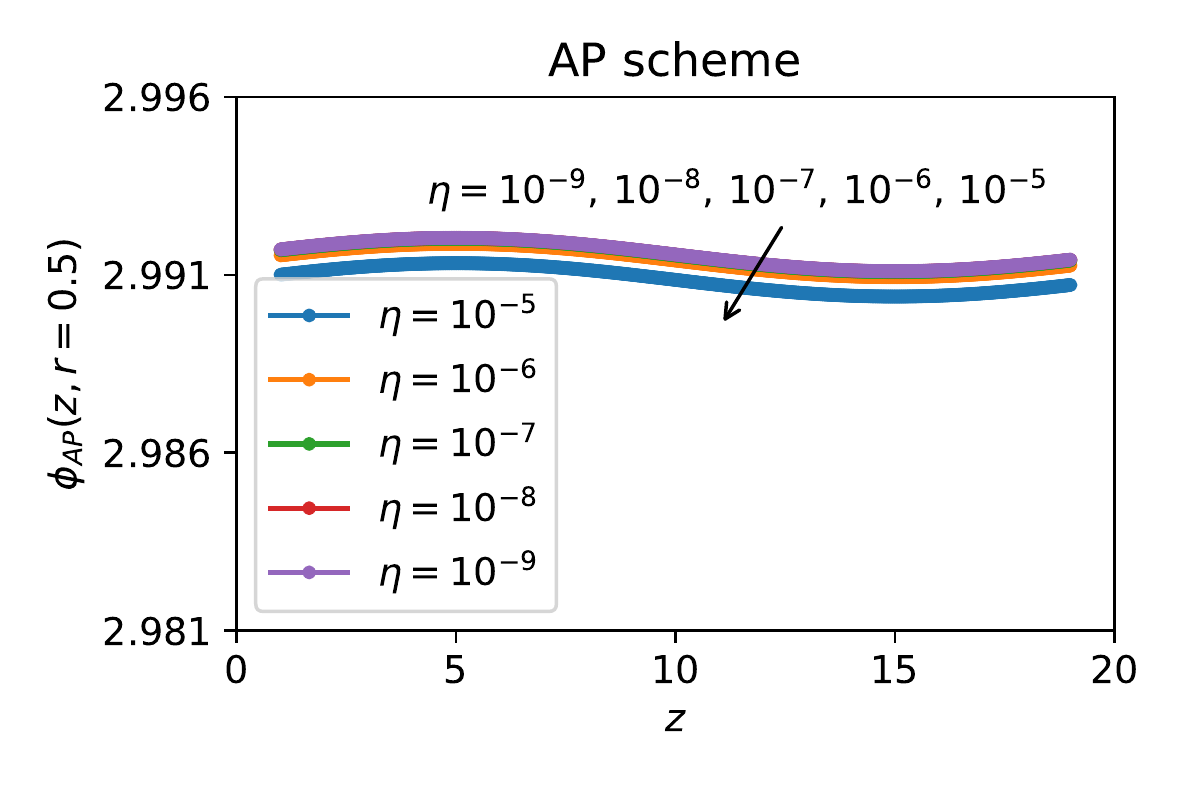}
	\includegraphics[scale=0.6]{\figdir/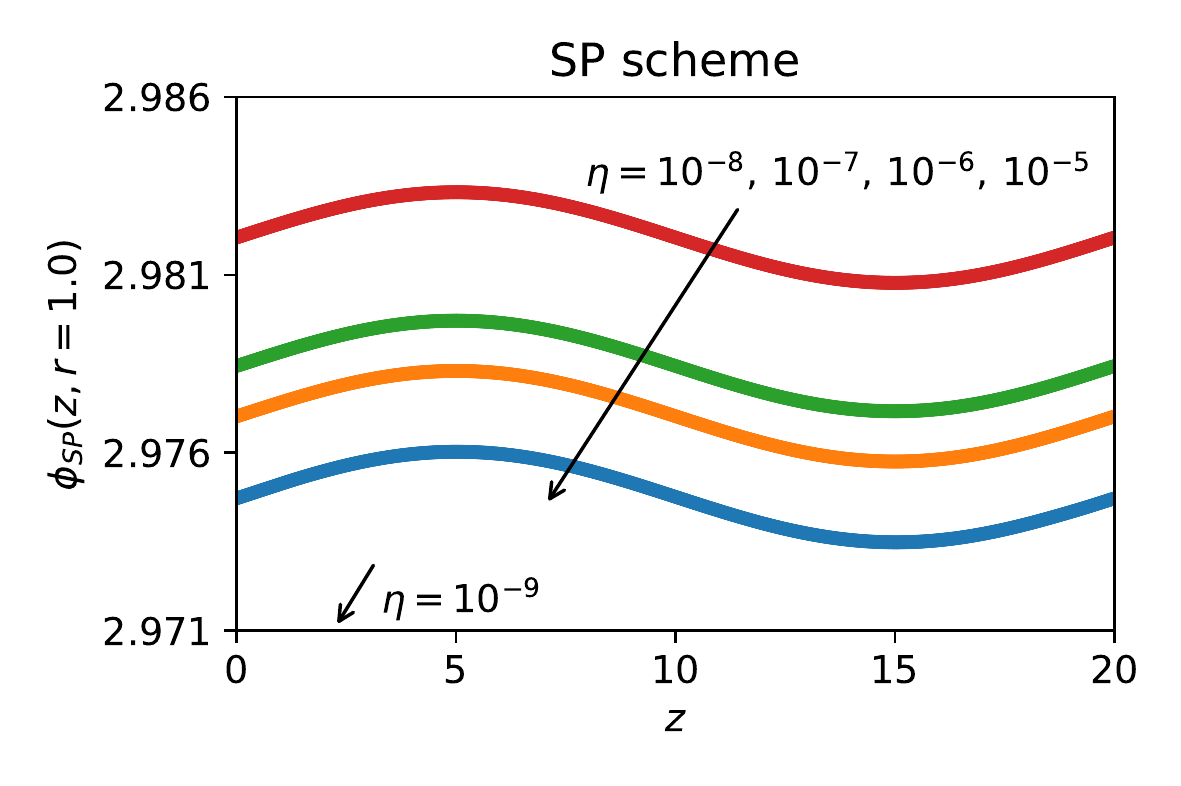}
	\includegraphics[scale=0.6]{\figdir/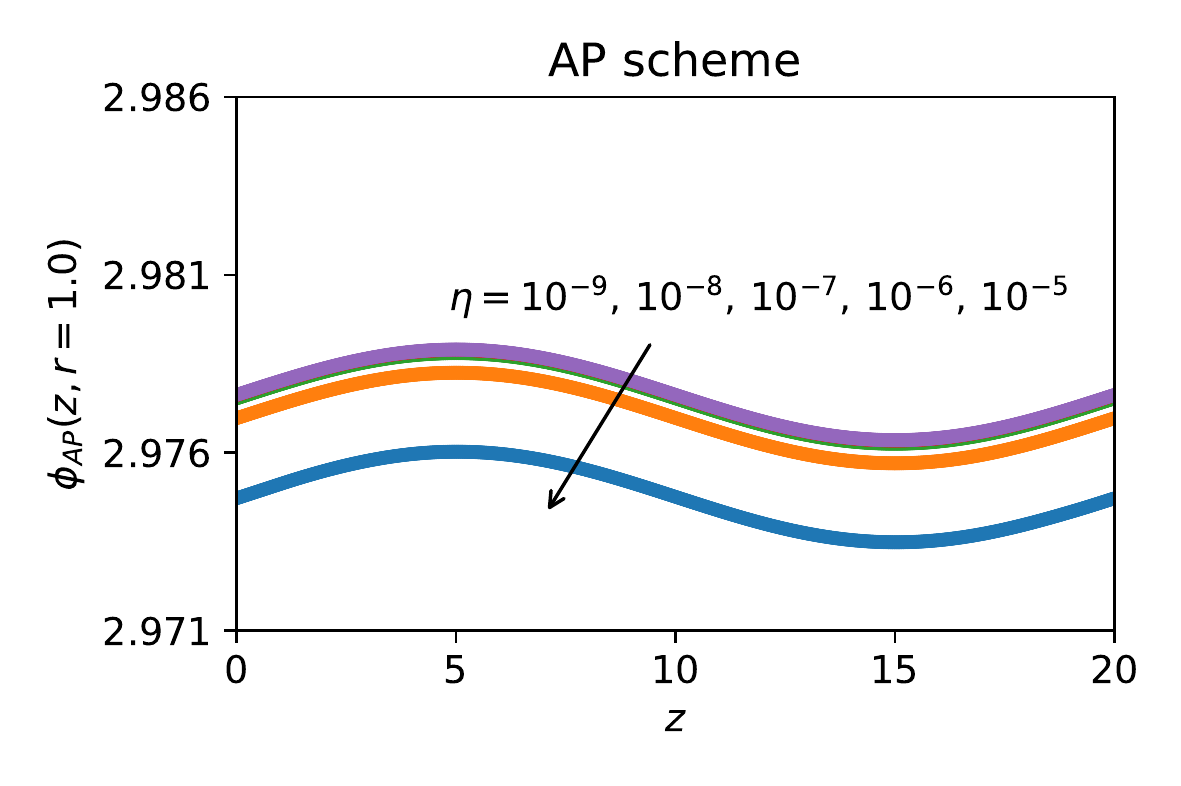}
	\includegraphics[scale=0.6]{\figdir/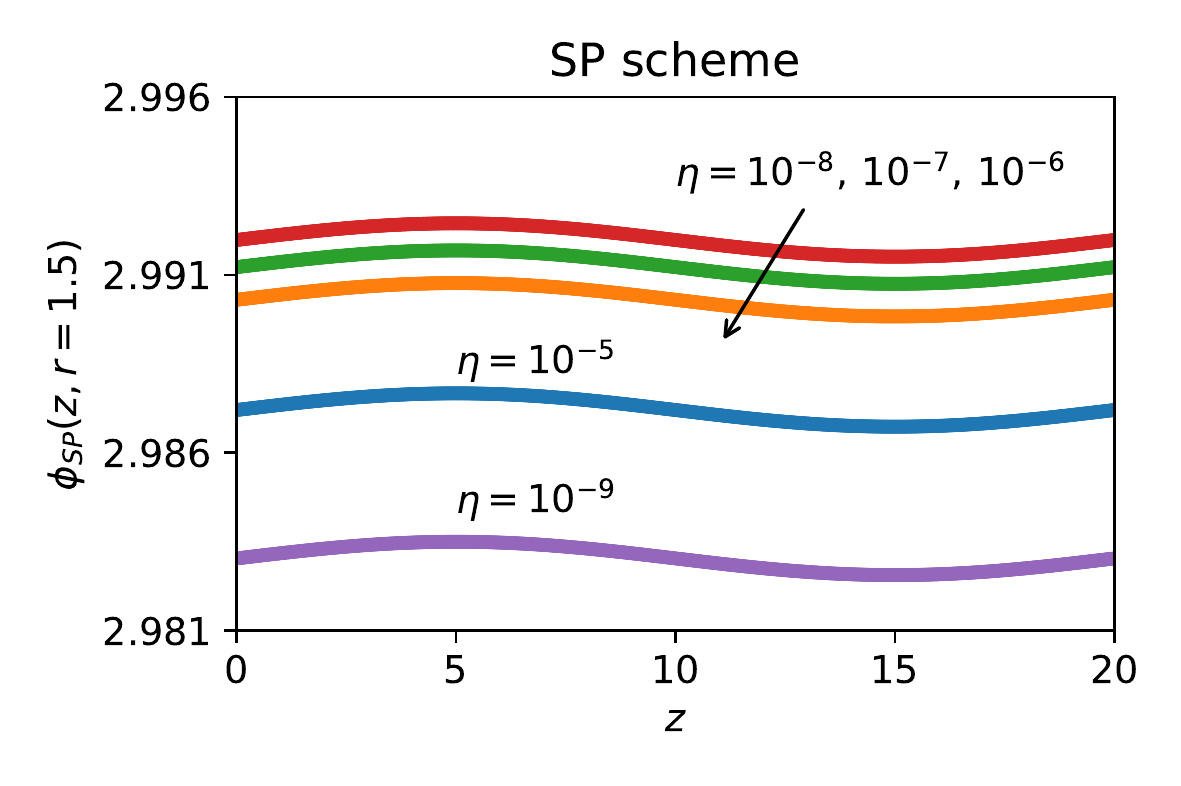}
	\includegraphics[scale=0.6]{\figdir/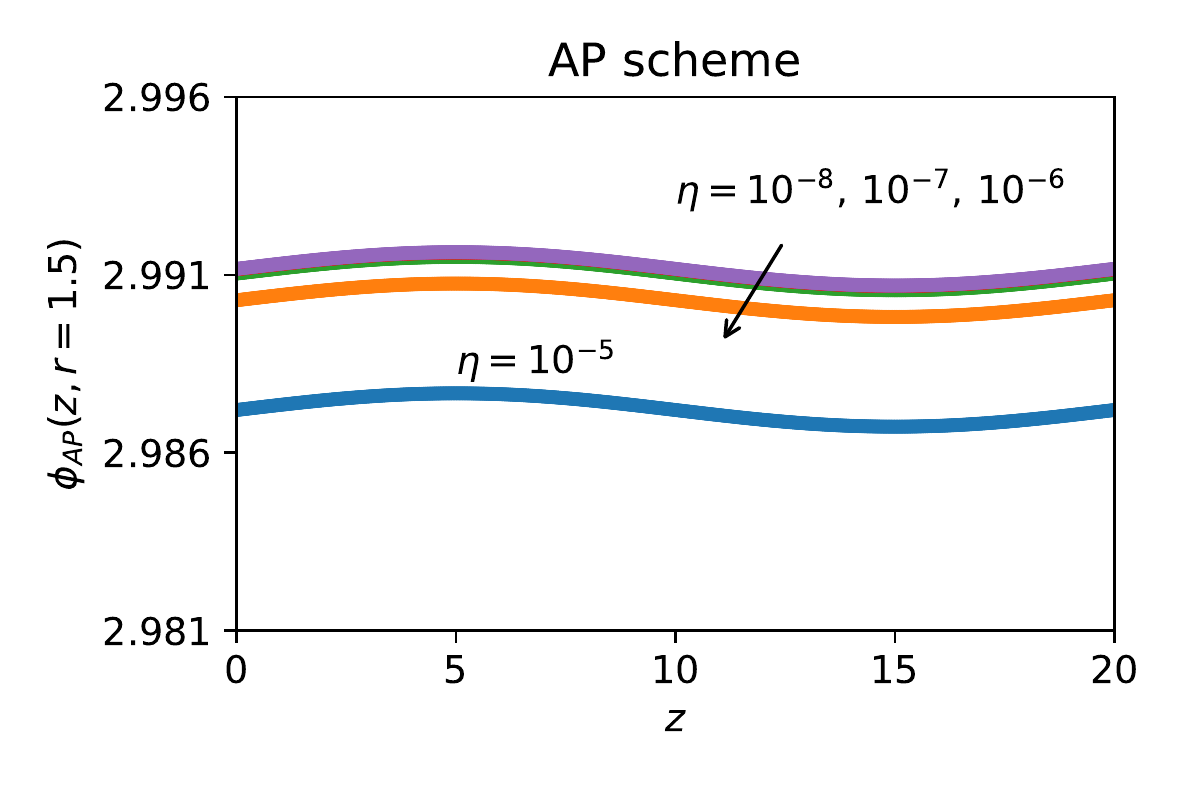}
	\caption{\textit{Case (P).} Comparison of the profiles of the solution $\phi$ as functions of $z$ at three different values of the radial coordinate: $r=0.5$ (top), $r=1$ (middle), $r=1.5$ (bottom) obtained with the {\SPeta } scheme (left) and with {\APeta } scheme (right) for various values of the parameter $\eta$ ($\eta=10^{-5}$, $10^{-6}$, $10^{-7}$, $10^{-8}$, $10^{-9}$). \label{fig:caseP_plotz} }
\end{figure}

\begin{figure}
	\includegraphics[scale=0.6]{\figdir/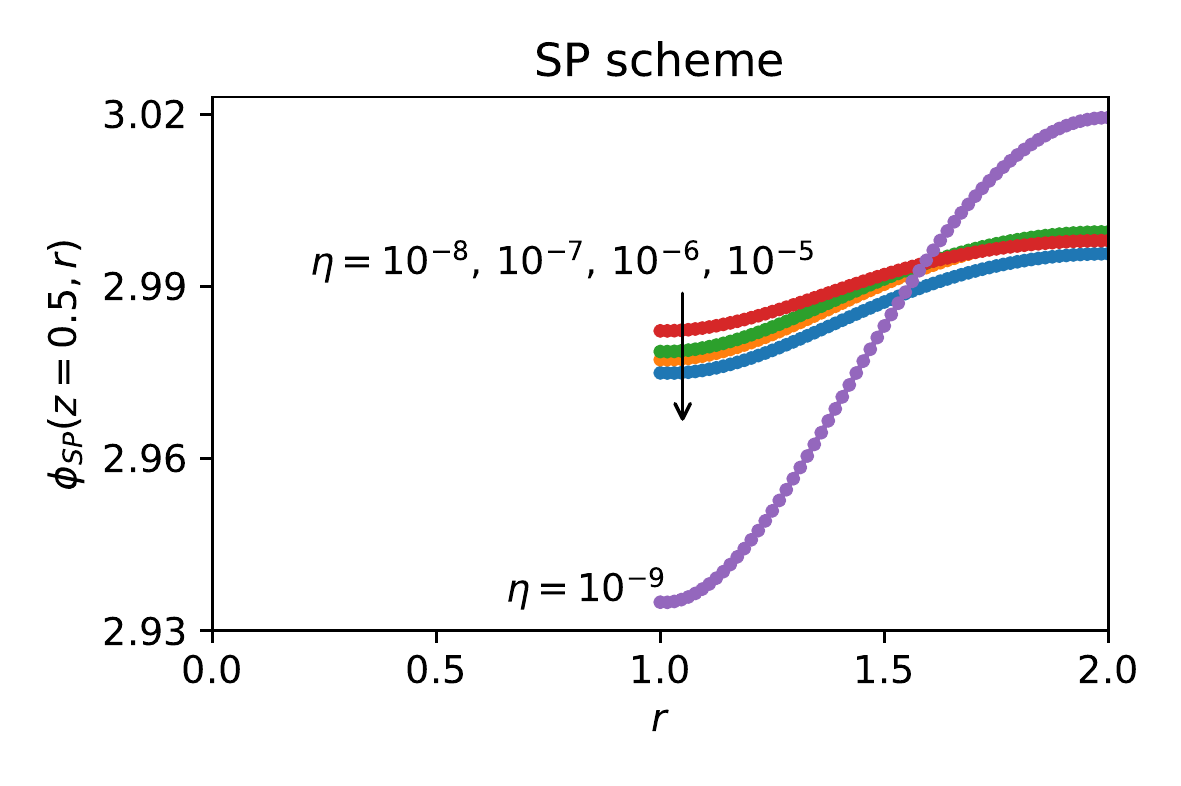}
	\includegraphics[scale=0.6]{\figdir/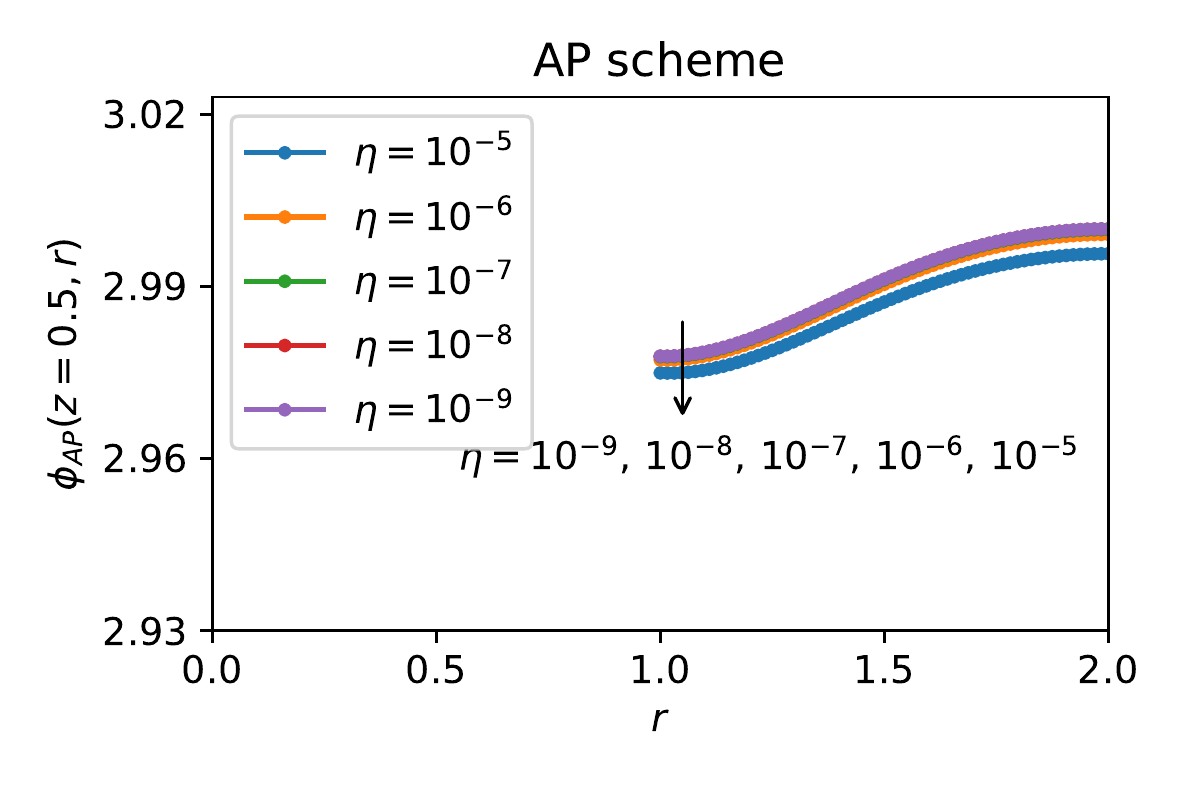}
	\includegraphics[scale=0.6]{\figdir/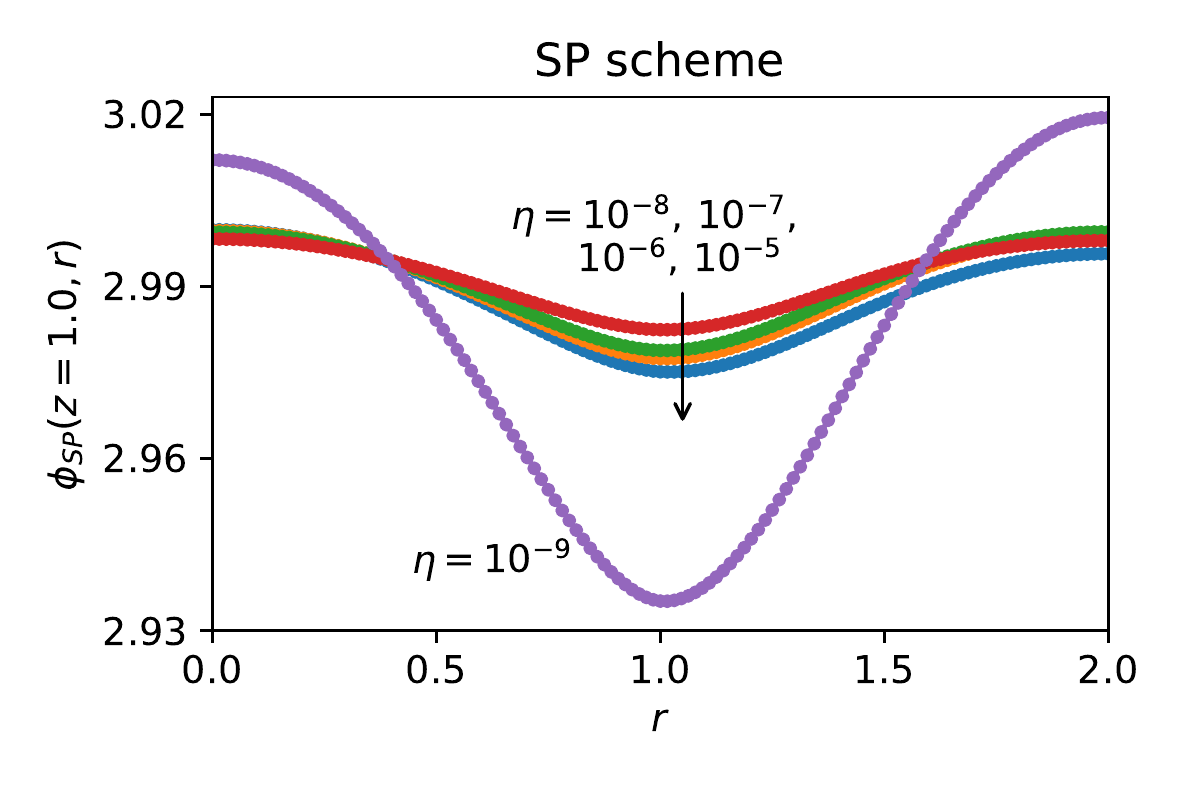}
	\includegraphics[scale=0.6]{\figdir/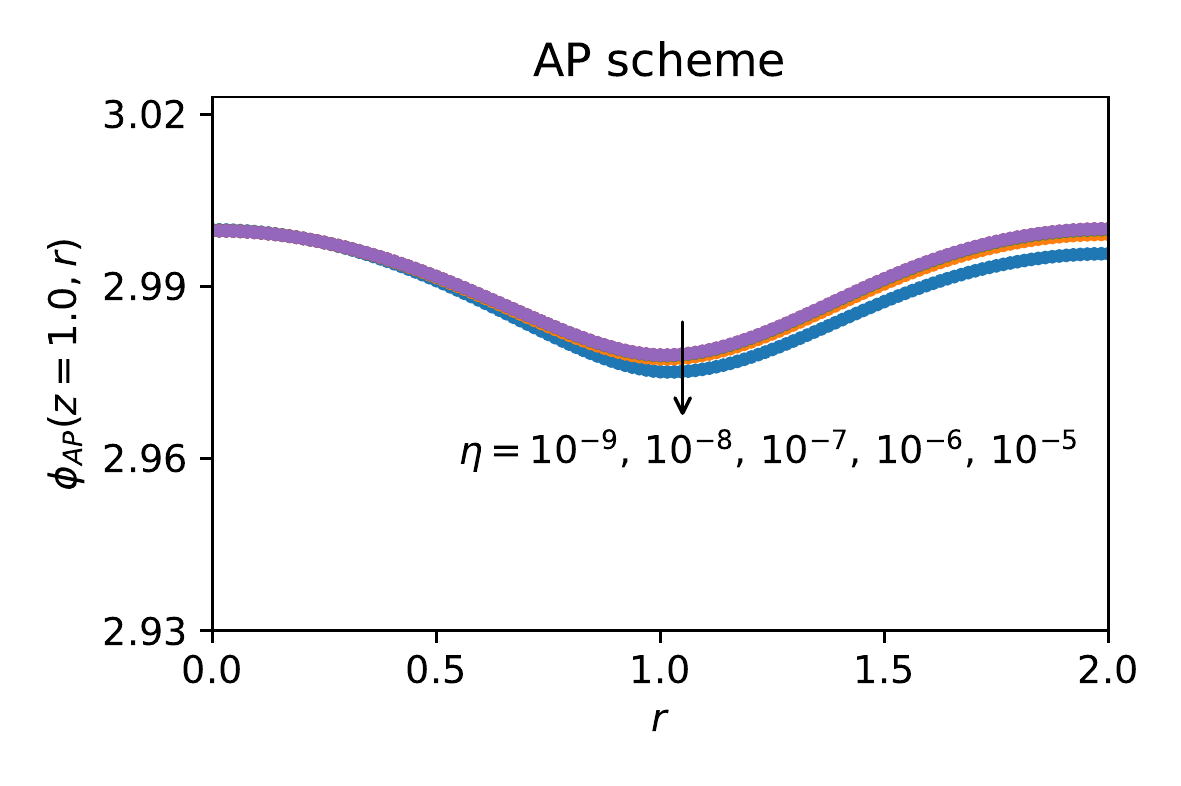}
	\includegraphics[scale=0.6]{\figdir/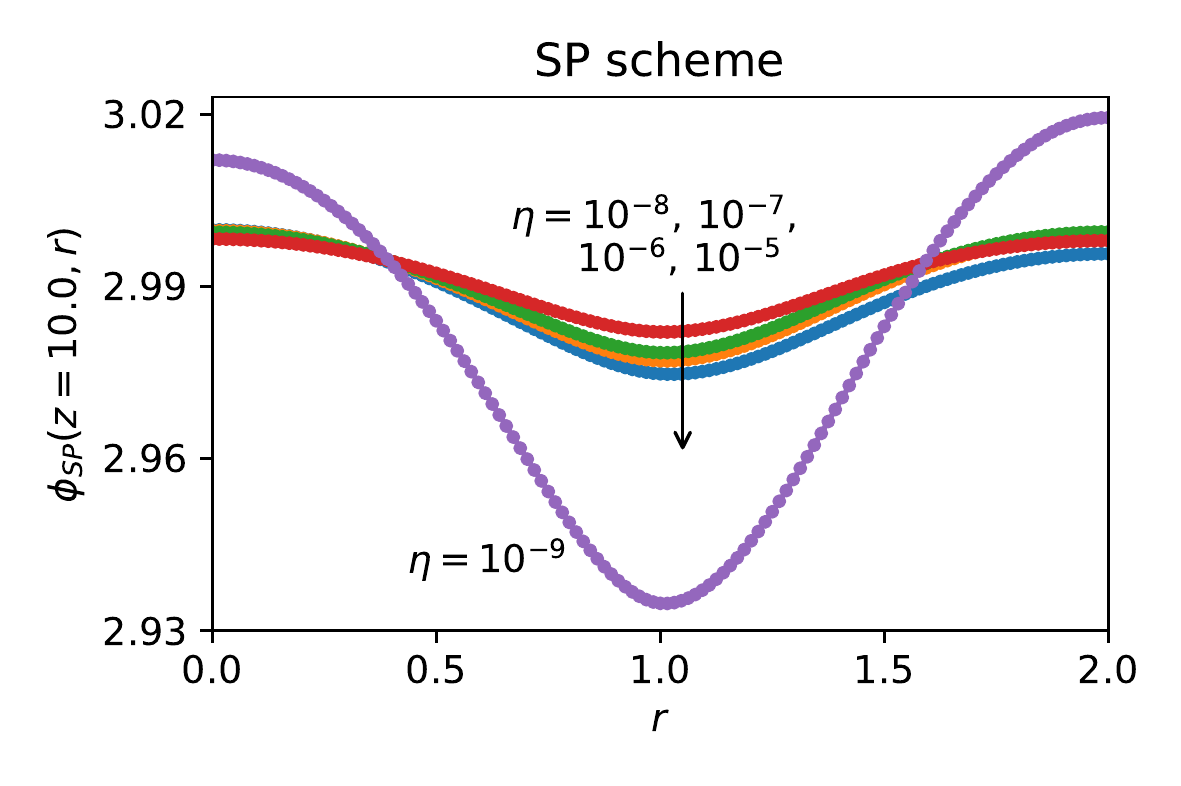}
	\includegraphics[scale=0.6]{\figdir/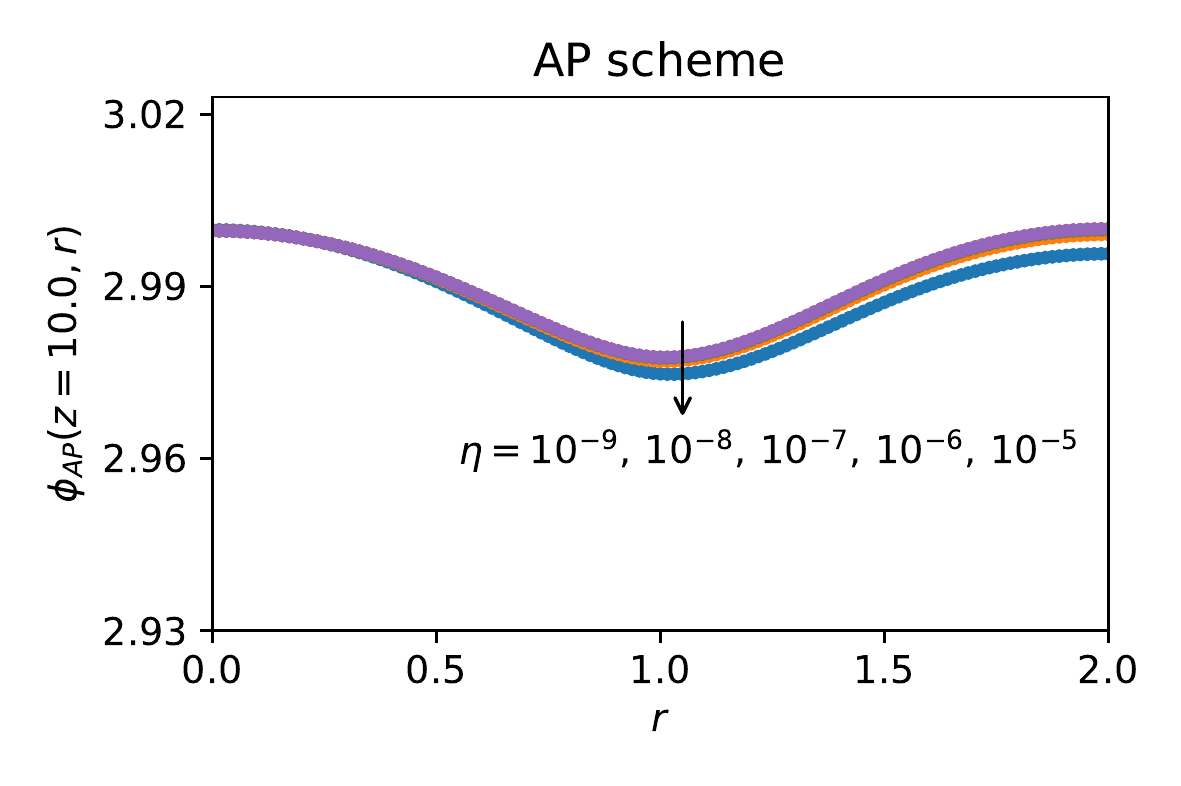}
	\caption{\textit{Case (P).} Comparison of the profiles of the solution $\phi$ as functions of $r$ at three different values of the longitudinal coordinate: $z=0.5$ (top), $z=1$ (middle), $z=10$ (bottom) obtained with the {\SPeta } scheme (left) and with {\APeta } scheme (right) for various values of the parameter $\eta$ ($\eta=10^{-5}$, $10^{-6}$, $10^{-7}$, $10^{-8}$, $10^{-9}$). \label{fig:caseP_plotr} }
\end{figure}

In Fig.~\ref{fig:caseP_plotz} and Fig.~\ref{fig:caseP_plotr}, we plotted the profiles of the solution $\phi^\eta$ as a function of $z$ at three different values of the radial coordinate, resp. as a function of $r$ at three different values of the longitudinal coordinate. The profiles obtained by means of the {\SPeta } scheme are compared to those obtained with the {\APeta } scheme, for several values of the parameter $\eta$. It is easily seen that as $\eta$ decreases, the profiles obtained with the {\APeta } scheme converge to the limit profile obtained with $\eta=0$. This correct, expected behavior is not observed in the profiles calculated by means of the {\SPeta } scheme. In this latter case, in fact, when $\eta$ is small enough ($\eta=10^{-9}$, in the case under investigation), the results are clearly not showing the expected trend. This behavior is naturally also pointed out by the $\eta$-evolution of the upper bound estimate of the relative error affecting the computed numerical solution of the discretized algebraic system associated to {\SPeta } (see Fig.~\ref{fig:caseP_cond}). The results obtained by means of the {\SPeta } scheme are therefore not reliable as the parameter $\eta$ approaches zero.\\
Looking in more details at Fig.~\ref{fig:caseP_plotz} and \ref{fig:caseP_plotr} and reminding the reduced problem \eqref{RR}, one can do a very nice observation. As stated in Section \ref{SEC2}, the solution to \eqref{RR} is not unique, as one can add an arbitrary $r$-dependent function $\psi(r)$ to one solution $\phi(r,z)$, in order to get another solution. The $z$-evolution of all these solutions is however well-defined. This specific feature can be now observed in Fig.~\ref{fig:caseP_plotz} and \ref{fig:caseP_plotr}. Indeed, for small $\eta$-values, instead of solving {\SPeta } the computer will solve the reduced problem \eqref{RR}, and will hence arbitrarily fix a function $\psi(r)$. Fig.~\ref{fig:caseP_plotz} shows precisely this behaviour, as the $z$-evolution is identical for both, {\SPeta } and {\APeta } schemes, however {\SPeta } has some difficulties below $\eta =10^{-8}$ to find the right constant, for fixed $r$. In  Fig.~ \ref{fig:caseP_plotr}, one notices immediately the wrong $r$-evolution for the {\SPeta }-scheme as $\eta$ gets smaller and smaller.

The computations shown in Fig.~\ref{fig:caseP_plotz} and Fig.~\ref{fig:caseP_plotr} have been carried out with $M_z=1\,280$, $M_r=129$, $M=156\,992$ ($M_U=156\,992$ for the {\SPeta } scheme; $M_U=313\,855$ for the {\APeta } scheme).\\

\smallskip 

\begin{figure}
	\includegraphics[scale=0.6]{\figdir/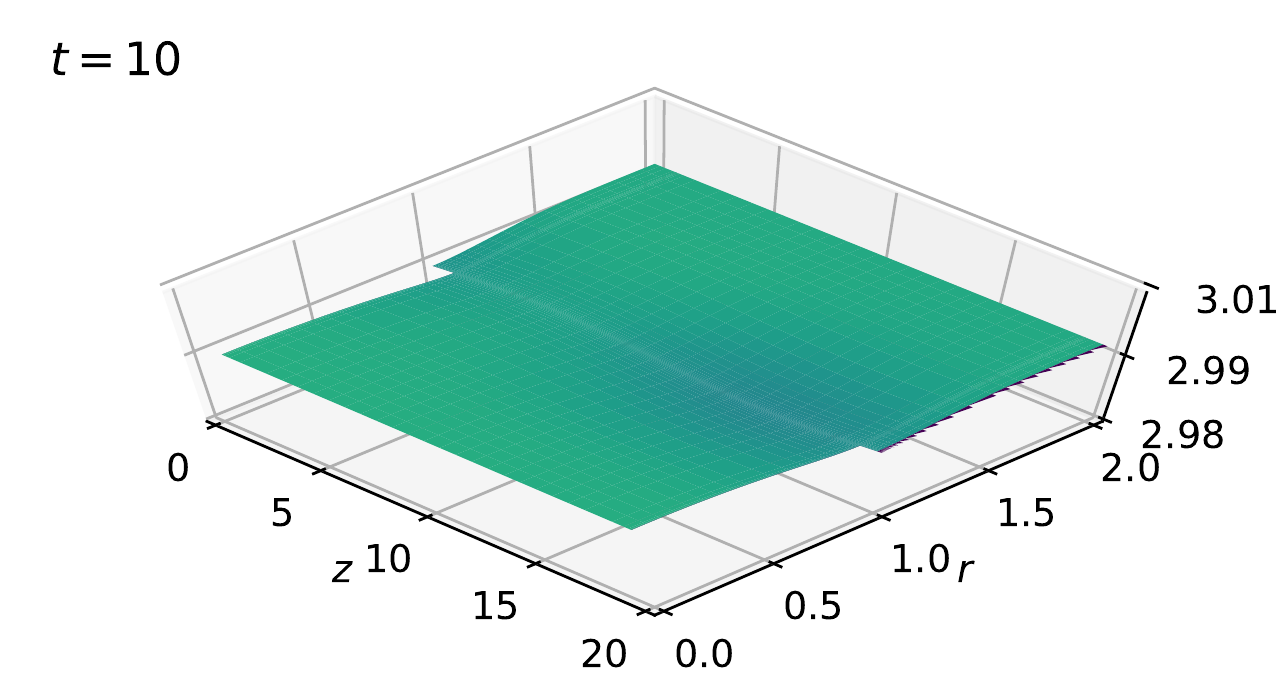}%
	\includegraphics[scale=0.6]{\figdir/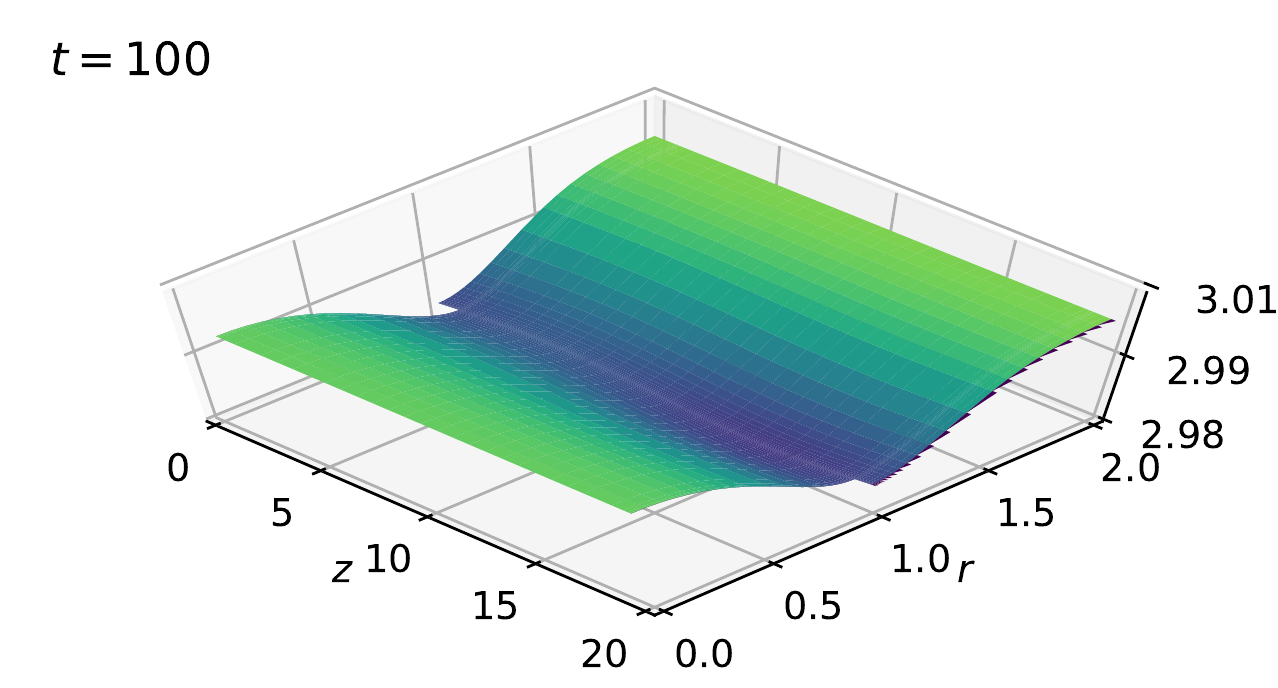}
	\includegraphics[scale=0.6]{\figdir/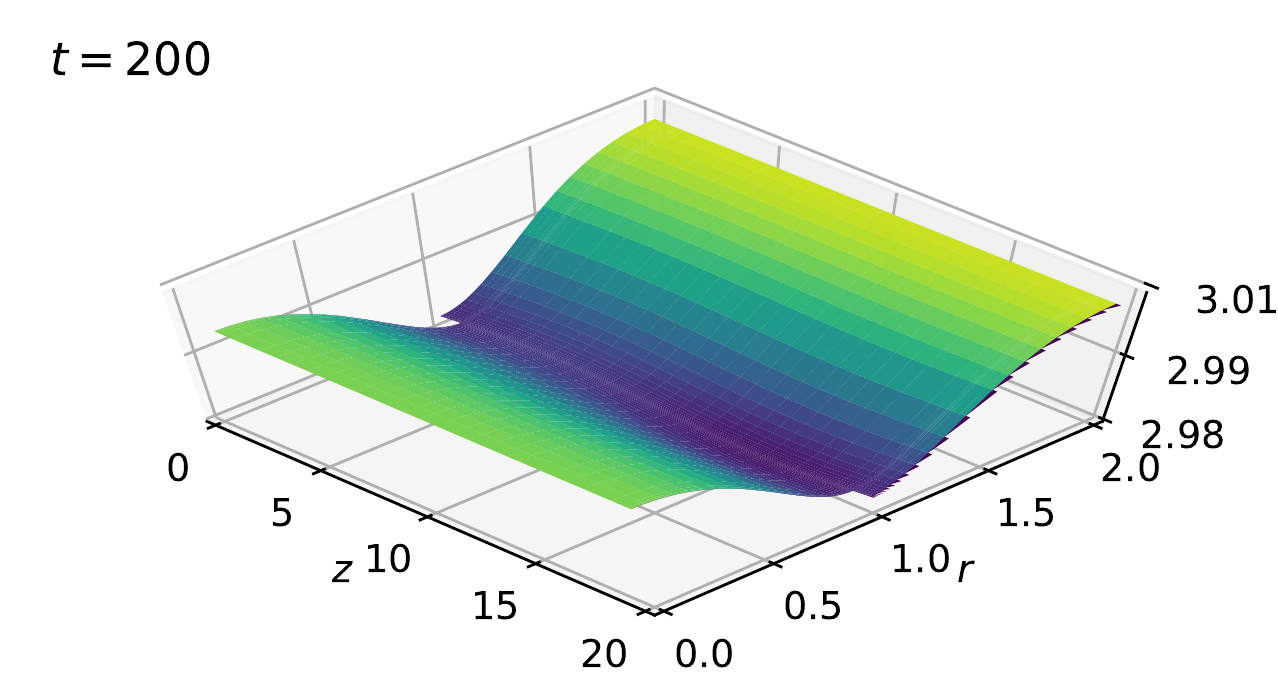}%
	\includegraphics[scale=0.6]{\figdir/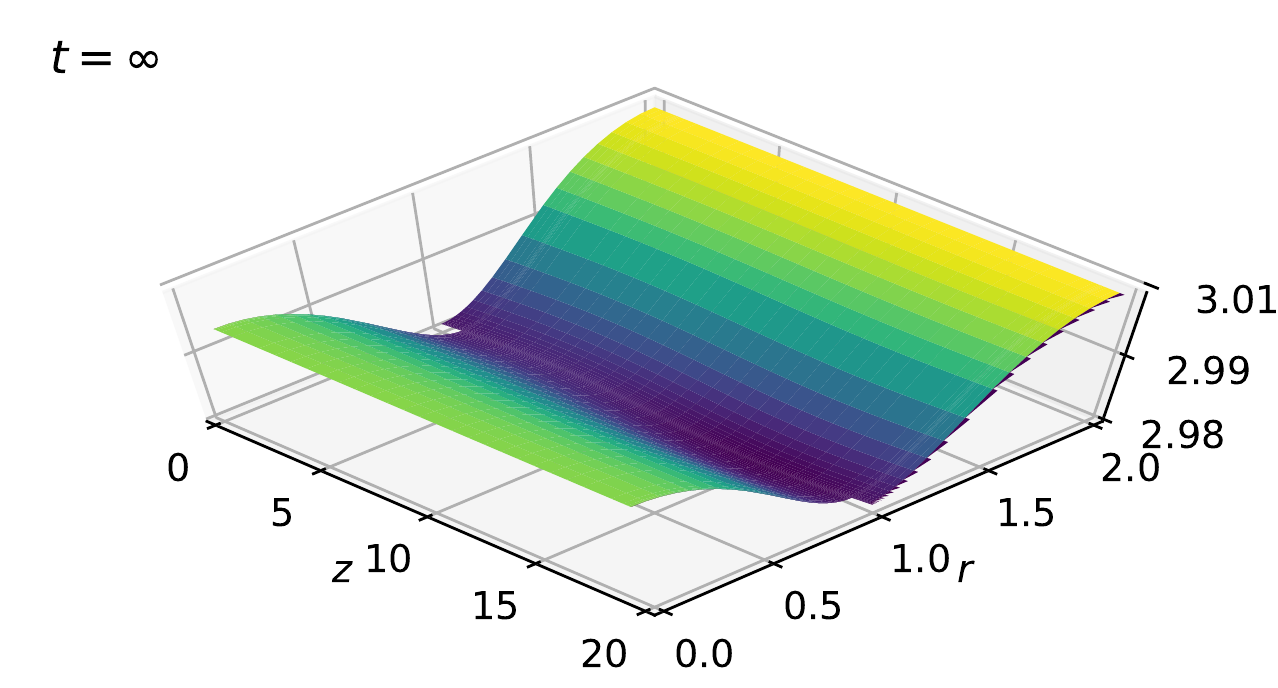}
	\caption{\textit{Case (P).} Evolution of the solution $\phi^\eta$, represented at four different time instants $t$. Here $\eta=10^{-12}$ and the results are obtained with the {\APeta } scheme. The initial condition $\phi_{in}(z,r)$ is given in Eq.~\eqref{eq:phi0}. \label{fig:caseP_evol_pcol} }
\end{figure}

In Fig.~\ref{fig:caseP_evol_pcol}, the evolution of the solution $\phi^\eta$ at four different time instants ($t=10$, $100$, $200$ and $t=\infty$) are represented. The initial data, as previously mentioned, is $\phi_{in}(z,r) = \Lambda$. The field $\phi^\eta$, obtained for $\eta=10^{-12}$ using the {\APeta } scheme, evolves towards the asymptotic solution represented in Fig.~\ref{fig:caseP_evol_pcol}(d) for large times. The last of the shown fields, corresponding to $t=\infty$, has been obtained with a time-independent version of the numerical code, and as such it represents the steady state solution. 

\begin{figure}
	\includegraphics[scale=0.6]{\figdir/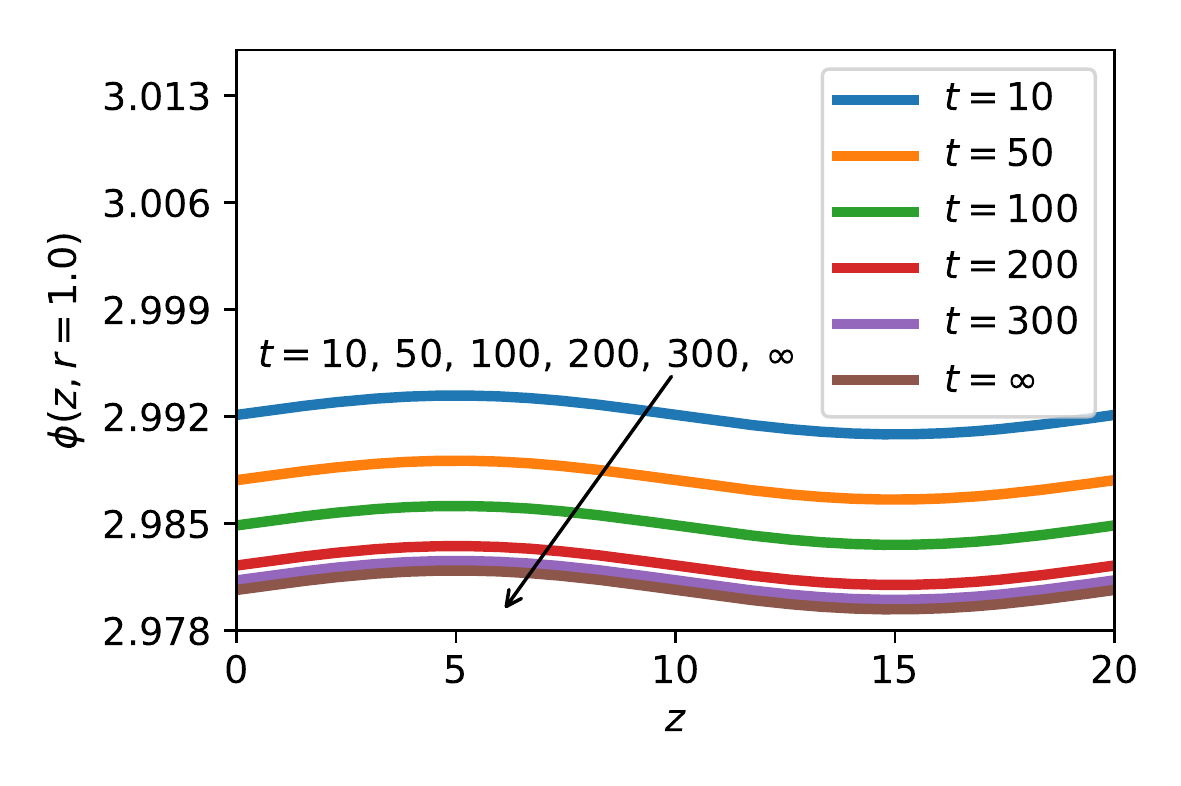}%
	\includegraphics[scale=0.6]{\figdir/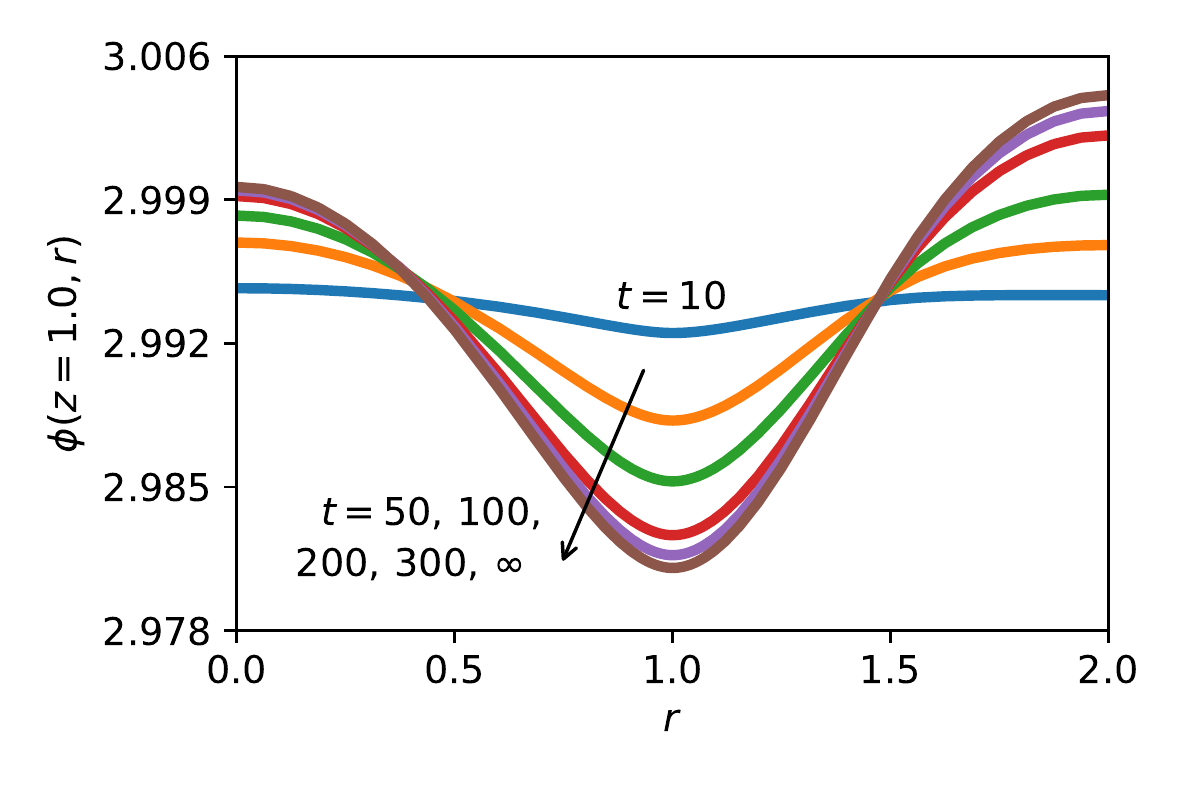}
	\caption{\textit{Case (P).} Longitudinal profiles for $r=1$ (left) and radial profiles for $z=1$ (right) of the solution $\phi^\eta$, represented at six different time instants $t$. Again $\eta=10^{-12}$ and the results are obtained with the {\APeta } scheme. The initial condition $\phi_{in}(z,r)$ is given in Eq.~\eqref{eq:phi0}. \label{fig:caseP_evol_plot} }
\end{figure}

In Fig.~\ref{fig:caseP_evol_plot} a selection of the longitudinal and radial profiles of the solution $\phi^\eta$ are plotted for several time instants, including those presented in Fig.~\ref{fig:caseP_evol_pcol}. From these profiles it is easy to observe that the solution converges to the steady-state solution labeled with $t=\infty$. Briefly, the AP-scheme seems to recover also very well the $t \rightarrow \infty$ asymptotics, and this for all values of $\eta \in [0,1]$.
The numerical solutions graphically shown in Fig.~\ref{fig:caseP_evol_pcol} and Fig.~\ref{fig:caseP_evol_plot} have been obtained with the {\APeta } scheme,  with $M_z=192$, $M_r=49$, $M=8\,280$ and $\Delta t = 1$. Grid refinement did not lead to appreciable changes in the results. 

\smallskip 

All the numerical results presented in this work have been obtained working in double precision floating-point arithmetics on a single node workstation, based on an Intel(R) Core(TM) i7-4770 CPU at 3.40GHz with hyper-threading enabled. A OpenMP-based parallelization has been adopted (when convenient) for the calculation of the linear system coefficients, and the MUMPS library has been compiled as to make use of OpenMP directives. In all cases, the computational time (\vv{wall time}) of each single computation ranges from few seconds to few minutes, such that no further form of software or hardware acceleration was needed for the purpose of the present study. 

\section{Conclusion}\label{SECC}
We introduced in this work an {\it Asymptotic-Preserving} numerical scheme for the resolution of the highly anisotropic vorticity equation arising in plasma modelling. Numerical simulations permitted to underline the advantages of this new scheme as compared to standard discretizations, used in present simulation codes. In particular our AP-scheme permits to choose the grid independent on the perturbation parameter $\eta$. This property can lead to an essential gain in memory and computational time, without loss of accuracy, for $\eta$-values below a certain threshold value. This threshold value can be situated in our test cases at approximately $\eta_* \in [10^{-8},10^{-6}]$. If the physical parameter $\eta$ is smaller than $\eta_*$, then our AP-methodology could be of important benefit for tokamak simulations. For larger $\eta$-values however, standard discretizations are to be preferred, as only one equation (for the unknown $\phi$) has to be solved, whereas our AP-scheme is constituted of two equations for $(\phi,q)$. Keep nevertheless in mind that the $\eta$-value can be varying in the simulation domain, taking different orders of magnitude in various parts of the domain. In such a case, the AP-formulation takes the advantage. Anyhow, our here presented AP-formulation has still to be extended to this situation of variable $\eta(x)$, as for the moment we treated only constant $\eta$-values. Another future step would be also to implement this new AP-method in the TOKAM3X code and validate its practical use. All this will be the aim of a future work.\\

\noindent {\bf Acknowledgments.} The authors would like first thank Patrick Tamain, Guido Ciraolo and Davide Galassi for fruitful discussions on this paper. Furthermore, we would  like to acknowledge support from the Italian National Group for Mathematical Physics (GNFM/INdAM). 
This work has been carried out within the framework of the EUROfusion Consortium and has received funding from the Euratom research and training programme 2014-2018 under grant agreement No 633053. The views and opinions expressed herein do not necessarily reflect those of the European Commission. 

\end{document}